\documentclass[
  11pt,
  a4paper,
  bibliography=totoc,
  numbers=noendperiod,
  parskip=half
]{scrartcl}

\usepackage[T1]{fontenc}
\usepackage[utf8]{inputenc}
\usepackage{lmodern}
\usepackage{microtype}
\usepackage{geometry}
\usepackage{xcolor}
\usepackage{graphicx}
\usepackage{subcaption}
\usepackage{booktabs}
\usepackage{longtable}
\usepackage{array}
\usepackage{rotating}
\usepackage{chngcntr}
\usepackage{enumitem}
\usepackage{siunitx}
\usepackage{listings}
\usepackage{tikz}
\usepackage{fontawesome5}
\usepackage{float}
\usepackage{csquotes}
\usepackage{needspace}
\usepackage[colorlinks=true,linkcolor=blue!50!black,citecolor=blue!50!black,urlcolor=blue!50!black]{hyperref}
\usepackage[nameinlink,noabbrev]{cleveref}
\usepackage[backend=biber,style=numeric-comp,sorting=none,maxbibnames=8]{biblatex}
\usetikzlibrary{arrows.meta,positioning,calc}

\definecolor{ebsdblue}{RGB}{0,84,159}
\definecolor{ebsdgreen}{RGB}{87,171,39}
\definecolor{ebsdorange}{RGB}{246,168,0}
\definecolor{codebg}{RGB}{247,247,247}

\graphicspath{{figures/}}

\lstdefinelanguage{yaml}{
  keywords={true,false,null,y,n},
  keywordstyle=\color{ebsdblue}\bfseries,
  basicstyle=\ttfamily\footnotesize,
  sensitive=false,
  comment=[l]{\#},
  morecomment=[l]{//},
  commentstyle=\color{gray},
  stringstyle=\color{ebsdgreen!60!black},
  morestring=[b]',
  morestring=[b]"
}
\newcommand{\orcidmark}[1]{\textsuperscript{\href{https://orcid.org/#1}{\textcolor{ebsdgreen!60!black}{\scriptsize iD}}}}
\newcommand{\yamllistingneed}{\Needspace{0.72\textheight}}
\newcommand{\personicon}{\textcolor{ebsdblue!85!black}{\faUser}}
\newcommand{\softwareicon}{\textcolor{black!70}{\faMicrochip}}

\title{EBSDmagus: Managing Multi-Stage Dynamical Electron Backscatter
  Diffraction Simulations}
\author{%
  Ulrich Kerzel\orcidmark{0000-0002-4939-6726}\textsuperscript{1}
  \qquad Lukas Berners\orcidmark{0009-0006-3418-0355}\textsuperscript{2}\\
  Sandra Korte-Kerzel\orcidmark{0000-0002-4143-5129}\textsuperscript{2}\\[0.75em]
  \small\textsuperscript{1}Faculty of Georesources and Materials Engineering,
  RWTH Aachen University, Aachen, Germany\\
  \small\textsuperscript{2}Institute of Physical Metallurgy and Materials Physics (IMM),\\
  \small RWTH Aachen University, Aachen, Germany
}
\date{Final author copy\\[0.3em]5 November 2026}
\usepackage[colorinlistoftodos,prependcaption,textsize=tiny]{todonotes} 

\usepackage{hyperref}

\begin{document}
\maketitle

\begin{abstract}
Dynamical electron backscatter diffraction (EBSD) simulations increasingly
support systematic studies of how material and experimental parameters affect
diffraction patterns. Investigations that combine composition and crystal
structure, including changes in atomic order, with accelerating voltage,
temperature, and many orientations or detector conditions can require large
sets of related calculations. Several packages provide dynamical EBSD
simulation; we use the open-source, scriptable EMsoft programs because they are
well suited to large studies on high-performance computing systems. An EMsoft
investigation nevertheless spans separate configuration, structure, scattering,
and orientation files and several dependent calculation stages. As variations
multiply, manual preparation becomes difficult to check, interruptions obscure
which results remain usable, and the origin of
individual patterns becomes laborious to reconstruct.

With EBSDmagus, researchers define fixed and varying parameters once. The
software prepares the required EMsoft calculations,
reuses compatible intermediate results, and checks that the generated files
represent the requested investigation before execution. After execution, it
checks the expected outputs and, following an interruption, resumes only
unresolved calculations while preserving completed work. A portable run record
links each output to its inputs and calculation history, providing the
provenance needed when the results are deposited as findable, accessible,
interoperable, and reusable (FAIR) data.

We assess this approach using a completed representative calculation, a larger
study prepared and checked before submission, two observed cluster
interruptions, and a controlled job cancellation. During recovery, EBSDmagus retained completed calculations
and resubmitted only incomplete work.
\end{abstract}


\begin{center}
  \centering
  \resizebox{0.96\linewidth}{!}{%
  \begin{tikzpicture}[
    gaBox/.style={
      draw=ebsdblue!65,
      fill=ebsdblue!4,
      rounded corners=3pt,
      minimum width=3.55cm,
      minimum height=2.75cm,
      align=center,
      font=\small
    },
    gaStage/.style={
      draw=ebsdblue!65,
      fill=ebsdblue!4,
      rounded corners=3pt,
      minimum width=3.55cm,
      minimum height=2.75cm,
      align=center,
      font=\small
    },
    gaHpc/.style={
      draw=ebsdblue!65,
      fill=ebsdblue!4,
      rounded corners=3pt,
      minimum width=3.55cm,
      minimum height=2.75cm,
      align=center,
      font=\small
    },
    gaDb/.style={
      draw=black!55,
      fill=black!4,
      rounded corners=3pt,
      minimum width=17.1cm,
      minimum height=1.0cm,
      align=center,
      font=\scriptsize
    },
    gaArrow/.style={-Latex, very thick, draw=ebsdblue!70},
    gaThinArrow/.style={-Latex, thick, draw=black!55},
    gaRecordArrow/.style={-Latex, thick, draw=black!45}
  ]
    \path[use as bounding box] (-1.98,-3.15) rectangle (15.63,1.62);

    \node[gaBox] (inputs) at (0,0) {};
    \node[font=\bfseries\small, align=center, text width=3.2cm] at (0,0.88)
      {Material and\\simulation inputs};
    \begin{scope}[xshift=-0.58cm,yshift=-0.28cm]
      \coordinate (a) at (-0.52,-0.50);
      \coordinate (b) at (0.14,-0.50);
      \coordinate (c) at (0.14,0.16);
      \coordinate (d) at (-0.52,0.16);
      \coordinate (e) at (-0.26,-0.24);
      \coordinate (f) at (0.40,-0.24);
      \coordinate (g) at (0.40,0.42);
      \coordinate (h) at (-0.26,0.42);
      \draw[ebsdblue!80, thick] (a) -- (b) -- (c) -- (d) -- cycle;
      \draw[ebsdblue!80, thick] (e) -- (f) -- (g) -- (h) -- cycle;
      \draw[ebsdblue!80, thick] (a) -- (e);
      \draw[ebsdblue!80, thick] (b) -- (f);
      \draw[ebsdblue!80, thick] (c) -- (g);
      \draw[ebsdblue!80, thick] (d) -- (h);
      \foreach \p in {a,b,c,d,e,f,g,h}
        \fill[ebsdblue!75] (\p) circle (0.085);
    \end{scope}
    \node[font=\scriptsize, align=left, anchor=west, text width=1.25cm]
      at (0.35,-0.18) {15/30~kV\\$T_1/T_2$\\Euler\\angles};

    \node[gaStage] (stages) at (4.55,0) {};
    \node[font=\bfseries\small, align=center, text width=3.2cm] at (4.55,0.88)
      {EBSDmagus\\+ EMsoft};
    \node[draw=black!55, fill=white, rounded corners=2pt, minimum width=0.82cm, minimum height=0.45cm, font=\scriptsize] (mc) at (3.40,0.18) {MC};
    \node[draw=black!55, fill=white, rounded corners=2pt, minimum width=0.82cm, minimum height=0.45cm, font=\scriptsize] (mp) at (4.55,0.18) {MP};
    \node[draw=black!55, fill=white, rounded corners=2pt, minimum width=0.82cm, minimum height=0.45cm, font=\scriptsize] (sp) at (5.70,0.18) {SP};
    \draw[gaThinArrow] (mc) -- (mp);
    \draw[gaThinArrow] (mp) -- (sp);

    \node[gaHpc] (hpc) at (9.1,0) {};
    \node[font=\bfseries\small, align=center, text width=3.2cm] at (9.1,0.88)
      {Many cluster jobs};
    \foreach \x in {8.1,8.72,9.34,9.96} {
      \foreach \y in {0.12,-0.28,-0.68} {
        \draw[ebsdblue!65, fill=white, rounded corners=1pt] (\x,\y) rectangle ++(0.42,0.25);
        \draw[ebsdblue!65] (\x+0.05,\y+0.07) -- ++(0.31,0);
      }
    }

    \node[gaBox] (library) at (13.65,0) {};
    \node[font=\bfseries\small, align=center, text width=3.2cm] at (13.65,0.88)
      {Pattern library};
    \node[inner sep=0pt] at (13.65,-0.04)
      {\includegraphics[width=2.55cm,trim=55bp 15bp 50bp 65bp,clip]{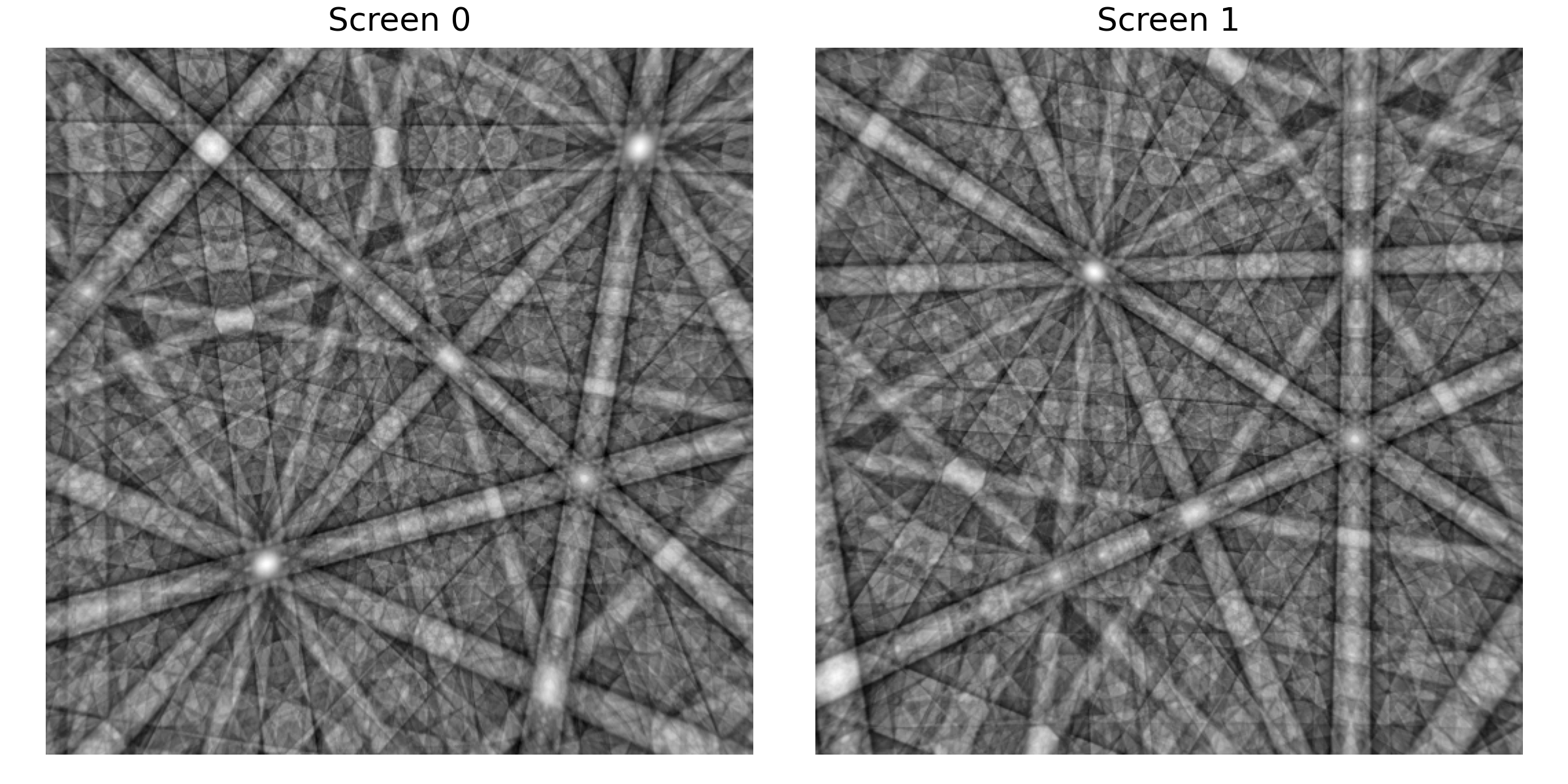}};

    \node[gaDb] (db) at (6.825,-2.40)
      {\textbf{Complete run record:} recovery, provenance and FAIR data};

    \draw[gaArrow] (inputs.east) -- (stages.west);
    \draw[gaArrow] (stages.east) -- (hpc.west);
    \draw[gaArrow] (hpc.east) -- (library.west);
    \draw[gaRecordArrow] (inputs.south) -- ($(db.north west)+(1.775,0)$);
    \draw[gaRecordArrow] (stages.south) -- ($(db.north west)+(6.325,0)$);
    \draw[gaRecordArrow] (hpc.south) -- ($(db.north west)+(10.875,0)$);
    \draw[gaRecordArrow] (library.south) -- ($(db.north west)+(15.425,0)$);
  \end{tikzpicture}%
  }
\end{center}

\section{Introduction}
\label{sec:introduction}

Dynamical electron backscatter diffraction (EBSD) simulations allow researchers
to examine how diffraction patterns respond to changes in a material and in the
experimental conditions. They can, for example, show how the intensity and
shape of Kikuchi bands vary with composition, sublattice ordering, defect
structure, thermal vibration, or beam conditions, while orientation and
detector geometry are varied separately
\cite{Callahan2013DynamicalEBSDPartI,Winkelmann2007ManyBeamEBSD,
Zhu2020DefectInteractionVolume}. Indexing is one application: simulated patterns
provide references for dictionary indexing \cite{Chen2015DictionaryIndexing},
while spherical indexing compares spherical-harmonic representations of a
simulated master pattern and an experimental pattern
\cite{Lenthe2019SphericalIndexing}.

Studies that combine several material and experimental variables may require
separate structures for the material states under investigation, with many
diffraction settings, orientations, and detector projections for each structure
and beam condition. The resulting calculations support both the study
of pattern formation and quantitative analysis, but managing them involves more
than submitting many independent jobs.

Several software packages provide dynamical EBSD simulations through different
interfaces and calculation approaches
\cite{BrukerESPRITDynamicS,OxfordAZtecCrystal,EDAXOIMMatrix}.
We use EMsoft \cite{Singh2017EMsoft} because its open-source command-line
programs can be scripted and distributed across high-performance computing
(HPC) systems, making it particularly suitable for large parameter studies. In
EMsoft, an EBSD
pattern is produced by a sequence of linked programs, each with its own
configuration and supporting files. A parameter study therefore creates many
closely related inputs, while intermediate results should be reused wherever
the material and simulation settings allow. \Cref{sec:dyn-ebsd-emsoft}
describes this calculation sequence in more detail.

On a shared HPC cluster, the process controlling the calculation sequence,
individual worker jobs, and the filesystem can fail independently. The
simulated patterns must nevertheless remain attributable to the material
description, parameter choices, and earlier calculations from which they
arose. The findable, accessible,
interoperable, and reusable (FAIR) principles extend this requirement beyond
the immediate run \cite{Wilkinson2016FAIR,Wilkinson2025FAIRComputationalWorkflows}:
the information needed to inspect or reuse a study should be recorded while the
calculations are prepared and executed, rather than reconstructed for
publication afterwards.

We developed EBSDmagus to keep one study definition intact through preparation,
cluster execution, recovery, and archiving. Common settings and parameter
sweeps are defined once in an editable YAML file. EBSDmagus resolves this
description into the required EMsoft calculations and generates a consistent
set of input files. It checks the prepared study before
cluster time is requested, while a portable database retains the run
history and output checks. After an interruption, EBSDmagus compares the
available records, keeps completed work that passes the defined checks, and
identifies what remains to be run. The complete run record can later accompany
the simulated patterns in a repository or institutional research data
management (RDM) system.

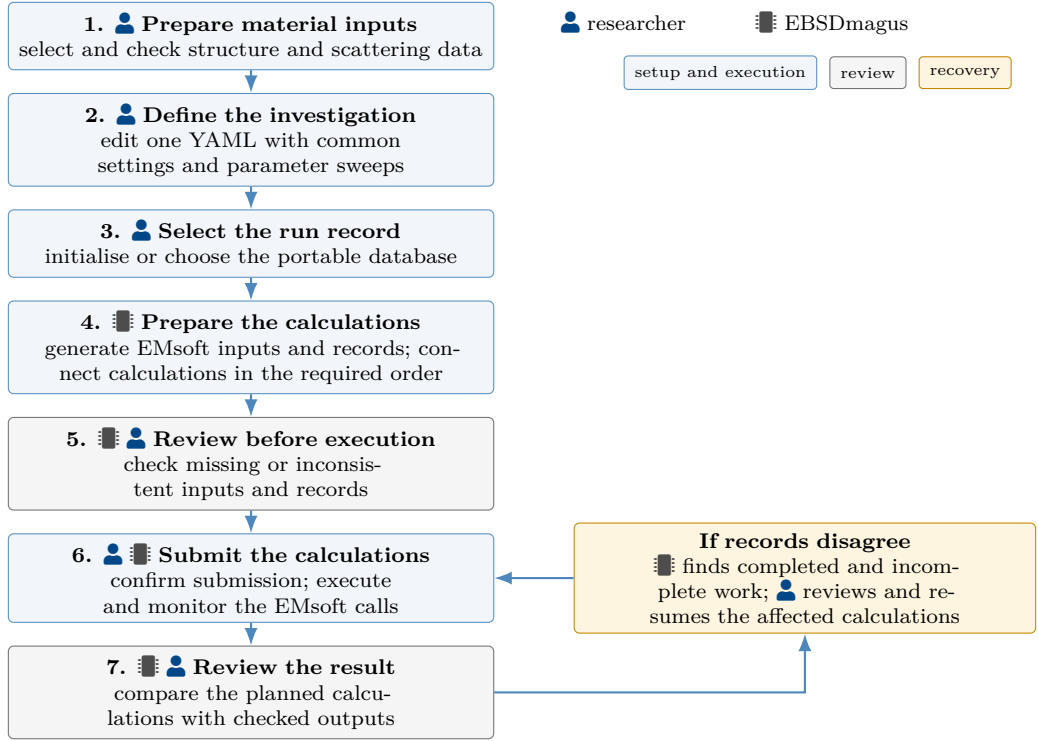
\begin{figure}[!htbp]
  \centering
  \begin{tikzpicture}[
    route/.style={
      draw=ebsdblue!65,
      fill=ebsdblue!5,
      rounded corners=2pt,
      align=center,
      minimum height=0.85cm,
      text width=0.39\linewidth,
      font=\scriptsize
    },
    check/.style={
      draw=black!55,
      fill=black!4,
      rounded corners=2pt,
      align=center,
      minimum height=0.85cm,
      text width=0.39\linewidth,
      font=\scriptsize
    },
    recover/.style={
      draw=ebsdorange!80!black,
      fill=ebsdorange!12,
      rounded corners=2pt,
      align=center,
      minimum height=1.1cm,
      text width=0.37\linewidth,
      font=\scriptsize
    },
    routeLegend/.style={
      rounded corners=1.5pt,
      minimum height=0.42cm,
      inner xsep=4pt,
      font=\tiny
    },
    arrow/.style={-Latex, thick, draw=ebsdblue!70}
  ]
    \node[route] (inputs) {\textbf{1. \personicon\ Prepare material inputs}\\select and check structure and scattering data};
    \node[route, below=0.30cm of inputs] (yaml) {\textbf{2. \personicon\ Define the investigation}\\edit one YAML with common settings and parameter sweeps};
    \node[route, below=0.30cm of yaml] (db) {\textbf{3. \personicon\ Select the run record}\\initialise or choose the portable database};
    \node[route, below=0.30cm of db] (prepare) {\textbf{4. \softwareicon\ Prepare the calculations}\\generate EMsoft inputs and records; connect calculations in the required order};
    \node[check, below=0.30cm of prepare] (preflight) {\textbf{5. \softwareicon\ \personicon\ Review before execution}\\check missing or inconsistent inputs and records};
    \node[route, below=0.30cm of preflight] (submit) {\textbf{6. \personicon\ \softwareicon\ Submit the calculations}\\confirm submission; execute and monitor the EMsoft calls};
    \node[check, below=0.30cm of submit] (postflight) {\textbf{7. \softwareicon\ \personicon\ Review the result}\\compare the planned calculations with checked outputs};
    \node[recover, right=1.05cm of submit] (recovery) {\textbf{If records disagree}\\\softwareicon\ finds completed and incomplete work; \personicon\ reviews and resumes the affected calculations};

    \node[anchor=west,font=\scriptsize] (legendPerson) at ($(inputs.east)+(0.75,0.18)$)
      {\personicon\ researcher};
    \node[right=0.65cm of legendPerson,font=\scriptsize] (legendSoftware)
      {\softwareicon\ EBSDmagus};
    \node[routeLegend,draw=ebsdblue!65,fill=ebsdblue!5,
      below=0.16cm of legendPerson,anchor=north west] (legendRoute)
      {setup and execution};
    \node[routeLegend,draw=black!55,fill=black!4,
      right=0.16cm of legendRoute] (legendCheck) {review};
    \node[routeLegend,draw=ebsdorange!80!black,fill=ebsdorange!12,
      right=0.16cm of legendCheck] {recovery};

    \draw[arrow] (inputs) -- (yaml);
    \draw[arrow] (yaml) -- (db);
    \draw[arrow] (db) -- (prepare);
    \draw[arrow] (prepare) -- (preflight);
    \draw[arrow] (preflight) -- (submit);
    \draw[arrow] (submit) -- (postflight);
    \draw[arrow] (postflight.east) -- (recovery.south |- postflight.east) -- (recovery.south);
    \draw[arrow] (recovery.west) -- (submit.east);
  \end{tikzpicture}
  \caption{Sequence followed when using EBSDmagus. The symbols identify actions
  taken by the researcher and by EBSDmagus; blue denotes setup and execution,
  grey review, and amber recovery.}
  \label{fig:user-route}
\end{figure}

YNi$_2$, an ordered intermetallic related to ongoing work, serves as the
example throughout the paper. After placing EBSDmagus in the context of EBSD
simulation software, workflow systems, and RDM practice, we describe the method
and report the calculation and recovery tests. \Cref{fig:user-route} outlines
this route; the appendices and archived package contain the detailed
configurations, database tables, and run records.

\section{Software and Data Management for Dynamical EBSD Studies}
\label{sec:related-work}

Related dynamical EBSD calculations must remain connected to their material
inputs, execution history, and resulting files. Existing EBSD software,
workflow frameworks, and RDM practice address different parts of this
requirement.

\subsection{Dynamical EBSD and EMsoft}
\label{sec:dyn-ebsd-emsoft}
Dynamical EBSD simulations link crystal structure and electron scattering to
the patterns acquired with an EBSD detector
\cite{Callahan2013DynamicalEBSDPartI,Winkelmann2007ManyBeamEBSD,Schwartz2009EBSDMaterialsScience}.
Commercial EBSD software also includes dynamical simulation and analysis, for
example Bruker ESPRIT DynamicS \cite{BrukerESPRITDynamicS}, Oxford Instruments
AZtecCrystal \cite{OxfordAZtecCrystal}, and EDAX OIM Matrix
\cite{EDAXOIMMatrix}. Each package follows its own interface and calculation
sequence. We use EMsoft \cite{Singh2017EMsoft} because it provides open-source
electron-diffraction and image-simulation programs whose command-line
executables can be scripted across HPC resources. These executables read
namelist (NML) inputs and write Hierarchical Data Format 5 (HDF5) files
\cite{Folk2011HDF5}.

For EBSD, EMsoft first calculates electron transport by Monte Carlo (MC), then
uses this result together with the crystal and scattering description to
generate a master pattern (MP). A screen-pattern (SP) calculation projects the
matching transport and master-pattern information for the chosen orientations
and detector geometry. If two later calculations use the same material
description and upstream simulation settings, they can share the corresponding
MC or MP result.

EMsoft organises pattern generation in these three linked stages, while other
simulation packages may divide the calculation differently. The EMsoft distribution also
provides Python wrappers through pyEMsoft, while the developing EMsoftOO version
includes a lightweight editor for individual namelists and program calls
\cite{EMsoftGitHub,EMsoftOOGitHub}. EBSDmagus coordinates the preparation,
checking, execution history, and recovery when many such calls form one
investigation. Support for another simulator would require a scriptable,
non-interactive interface together with suitable input templates, calculation
relationships, and output checks. The implementation and tests reported here
focus on EMsoft.

\subsection{Workflow Frameworks for Dynamical EBSD Studies}
Before workflow software can coordinate a dynamical EBSD study, it needs a
description of the calculation route used by the selected simulator. For
EMsoft, that description specifies which settings belong to MC, MP, or SP,
which earlier result supplies a later calculation, when that result may be used
again, and which files must agree before interrupted work resumes. These
relationships concern the complete three-stage calculation sequence rather than
any single namelist or executable.

Several frameworks have established calculation management and provenance as
part of computational materials research, with provenance recording the link
between a result and the calculations that produced it. These include AiiDA
\cite{Pizzi2016AiiDA,Huber2020AiiDA}, FireWorks \cite{Jain2015FireWorks},
jobflow \cite{Rosen2024Jobflow}, atomate \cite{Mathew2017Atomate}, and pyiron
\cite{Janssen2019Pyiron}.
They provide broad mechanisms for defining dependent calculations, using local
or remote resources, and preserving their history.
Janssen et al. \cite{Janssen2025PythonWorkflowDefinition} proposed the Python
Workflow Definition to support the exchange of dependent calculation
descriptions among AiiDA, jobflow, and pyiron. Applying any of these frameworks
to dynamical EBSD would still require the rules of the selected simulation
software. For the EMsoft route used here, these include stage-specific input
generation, compatible parent selection, and the checks used after a cluster
interruption.

On a shared HPC system, the services required by a workflow framework determine
whether researchers can install and operate the framework under an ordinary
user account. AiiDA can be installed under such an account, but its full
process control uses RabbitMQ
and its documentation recommends PostgreSQL for production and high-throughput
work \cite{AiiDAQuickInstallationDocs}. FireWorks records workflows in a
MongoDB LaunchPad and describes an offline mode for compute nodes that cannot
reach the database directly
\cite{FireWorksInstallationDocs,FireWorksOfflineDocs}. Pyiron and jobflow can
operate with less external service infrastructure. A group that already uses
one of these platforms can employ it to connect several simulation codes.

EBSDmagus prepares the crystal descriptions, scattering inputs, and dependent
calculations required by EMsoft. Its checks on crystal descriptions,
Debye--Waller values, and Bethe parameters are described in
\cref{sec:material-description,sec:yaml-definition}. It uses Snakemake
\cite{Koster2012Snakemake,Molder2025SustainableSnakemake} to order the
calculations, Jinja \cite{PalletsJinjaDocs} to render their inputs, and SQLite
\cite{SQLiteAbout} to retain the calculation record in one file. These
components run under an ordinary cluster account without a separate database
service. \Cref{sec:methods-overview} describes how EBSDmagus combines them to
prepare related calculations, compare generated inputs with the requested
study, and identify completed work that can be retained after an interruption.

\subsection{Research Data Management and FAIR Data}
Research funders increasingly require data to be managed throughout a project
and made as open and reusable as the subject permits. Horizon Europe, for
example, requires a data management plan and FAIR management of generated or
collected research data \cite{EuropeanCommissionOpenScience}; the DFG expects
funded projects to describe how they will document, store, preserve, and enable
later use of their data \cite{DFGResearchData}. NFDI-MatWerk develops
corresponding FAIR data practice and infrastructure for experimental and
simulated materials data
\cite{Eberl2021NFDIMatWerk,NFDIMatWerkProject}. The findable, accessible,
interoperable, and reusable (FAIR) principles \cite{Wilkinson2016FAIR} also
apply to the computational procedures that produced a dataset
\cite{Wilkinson2025FAIRComputationalWorkflows}. For a simulation study, FAIR
practice requires a record of how the final outputs were produced, not only the
output files themselves.

For simulated EBSD data, HDF5 and the h5EBSD format
\cite{Jackson2014H5EBSD} demonstrate how data and substantial metadata can
travel together. An individual EMsoft file can likewise contain detailed
calculation information. Such a file does not, however, describe all planned
calculations, their cluster history, or decisions made after an interruption.
EBSDmagus records this information during preparation and execution.

The source YAML and generated inputs accompany a database that records the
dependent calculations, cluster history, checks, and output paths; reports and
selected outputs may be added for inspection. The resulting package is suitable
for deposit in a repository such as Zenodo \cite{ZenodoRepository} or for
transfer to an institutional RDM system, where identifiers, access conditions,
licences, and repository metadata can be added.

\section{Methods}
\label{sec:methods}

\subsection{Overview of EBSDmagus}
\label{sec:methods-overview}

\begin{figure}[H]
  \centering
  \begin{subfigure}[t]{\linewidth}
  \centering
  \resizebox{\linewidth}{!}{%
  \begin{tikzpicture}[
    overviewInput/.style={
      draw=ebsdblue!70,
      fill=ebsdblue!5,
      rounded corners=3pt,
      text width=3.15cm,
      minimum height=2.7cm,
      align=center,
      font=\small
    },
    overviewPrepare/.style={
      draw=ebsdblue!70,
      fill=ebsdblue!5,
      rounded corners=3pt,
      text width=5.75cm,
      minimum height=3.15cm,
      align=center,
      font=\small
    },
    overviewCheck/.style={
      draw=black!55,
      fill=black!4,
      rounded corners=3pt,
      text width=3.15cm,
      minimum height=2.7cm,
      align=center,
      font=\small
    },
    overviewRun/.style={
      draw=ebsdblue!70,
      fill=ebsdblue!5,
      rounded corners=3pt,
      text width=3.9cm,
      minimum height=2.35cm,
      align=center,
      font=\small
    },
    overviewOutput/.style={
      draw=ebsdblue!70,
      fill=ebsdblue!5,
      rounded corners=3pt,
      text width=3.9cm,
      minimum height=2.35cm,
      align=center,
      font=\small
    },
    overviewArchive/.style={
      draw=black!55,
      fill=black!4,
      rounded corners=3pt,
      text width=3.9cm,
      minimum height=2.35cm,
      align=center,
      font=\small
    },
    overviewArrow/.style={-Latex, very thick, draw=black!60}
  ]
    \node[overviewInput] (source) {\textbf{Researcher defines}\\[0.2em]
      structure and scattering inputs\\fixed values and sweeps in one YAML};
    \node[overviewPrepare, right=0.65cm of source] (prepare) {\textbf{EBSDmagus prepares}\\[0.25em]
      EMsoft inputs and parent calculations\\
      portable database file};
    \node[overviewCheck, right=0.65cm of prepare] (before) {\textbf{Before the cluster run}\\[0.2em]
      compare the YAML, generated inputs, calculation graph, and database};

    \node[overviewRun, below=0.75cm of before] (run) {\textbf{Cluster execution}\\[0.2em]
      Snakemake submits MC, MP, and SP jobs via the SLURM profile};
    \node[overviewOutput, left=0.65cm of run] (outputs) {\textbf{Each job records}\\[0.2em]
      HDF5 output and log\\completion and structural checks};
    \node[overviewArchive, left=0.65cm of outputs] (after) {\textbf{After the run}\\[0.2em]
      compare planned and completed work\\resume unfinished branches\\archive outputs and the run record};

    \draw[overviewArrow] (source) -- (prepare);
    \draw[overviewArrow] (prepare) -- (before);
    \draw[overviewArrow] (before) -- (run);
    \draw[overviewArrow] (run) -- (outputs);
    \draw[overviewArrow] (outputs) -- (after);
  \end{tikzpicture}}
  \caption{}
  \label{fig:method-components-flow}
  \end{subfigure}

  \vspace{0.7em}
  \begin{subfigure}[t]{0.88\linewidth}
  \centering
  \resizebox{\linewidth}{!}{%
  \begin{tikzpicture}[
    depMC/.style={draw=black!55, fill=black!4, rounded corners=2pt,
      minimum width=2.4cm, minimum height=1.05cm, align=center, font=\small},
    depMP/.style={draw=black!55, fill=black!4, rounded corners=2pt,
      minimum width=2.6cm, minimum height=1.05cm, align=center, font=\small},
    depSP/.style={draw=black!55, fill=black!4, rounded corners=2pt,
      minimum width=2.5cm, minimum height=0.9cm, align=center, font=\small},
    depArrow/.style={-Latex, very thick, draw=black!60}
  ]
    \node[depMC] (mc) {MC\\electron transport};
    \node[depMP, right=2.0cm of mc, yshift=1.2cm] (mpa) {MP A\\diffraction settings A};
    \node[depMP, right=2.0cm of mc, yshift=-1.2cm] (mpb) {MP B\\diffraction settings B};
    \node[depSP, right=1.7cm of mpa, yshift=0.55cm] (spa1) {SP A1};
    \node[depSP, right=1.7cm of mpa, yshift=-0.55cm] (spa2) {SP A2};
    \node[depSP, right=1.7cm of mpb, yshift=0.55cm] (spb1) {SP B1};
    \node[depSP, right=1.7cm of mpb, yshift=-0.55cm] (spb2) {SP B2};

    \draw[depArrow] (mc) -- (mpa);
    \draw[depArrow] (mc) -- (mpb);
    \draw[depArrow] (mpa) -- (spa1);
    \draw[depArrow] (mpa) -- (spa2);
    \draw[depArrow] (mpb) -- (spb1);
    \draw[depArrow] (mpb) -- (spb2);
  \end{tikzpicture}}
  \caption{}
  \label{fig:method-components-dependencies}
  \end{subfigure}

  \caption{Files and calculation relationships in EBSDmagus. (a) The study
  definition is expanded by EBSDmagus into inputs and run records for comparison
  before execution and after outputs return. Blue denotes preparation and
  execution; grey denotes review steps. (b) One compatible MC result supplies
  two independent MP branches, each supplying two SP calculations. Arrows
  indicate prerequisite results; the neutral boxes do not denote completion.}
  \label{fig:method-components}
\end{figure}
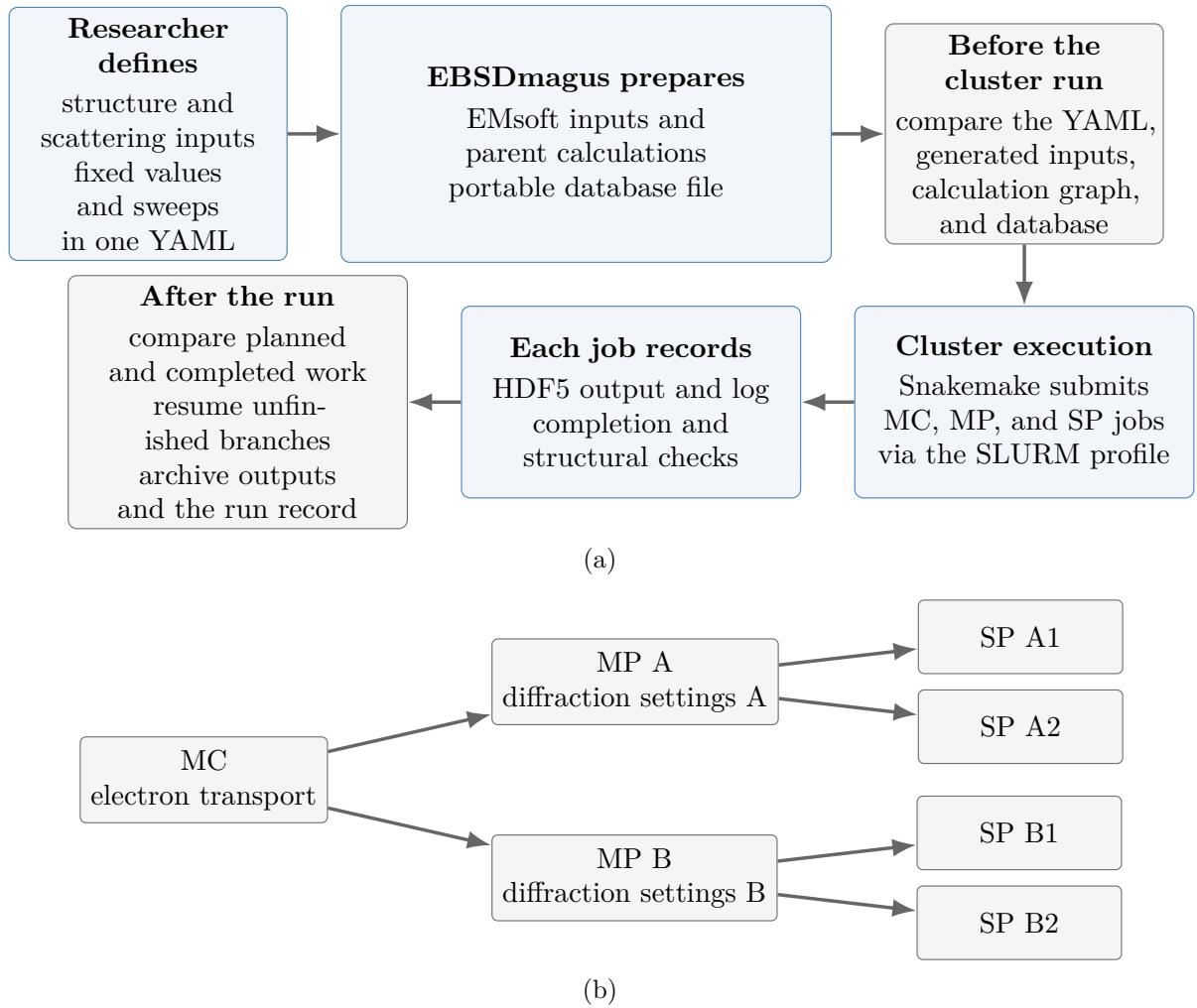

To prepare an EMsoft investigation, the researcher supplies the crystal
structure and scattering information and records the fixed settings and
parameter variations in one YAML file. EBSDmagus turns this description into
the individual EMsoft calls and their input files, checks them before
submission, and later compares the completed work with the original request.
\Cref{fig:method-components-flow} follows the files and records through these
steps; \cref{sec:material-description,sec:yaml-definition} describe the material
inputs and study settings.

The EMsoft calls cannot be made in an arbitrary order. A master-pattern
calculation requires the result of a compatible Monte Carlo calculation, and a
screen-pattern calculation requires the corresponding master-pattern output.
At the same time, two branches that do not depend on one another can run
independently. For example, one MC result may support two MP variants, each of
which may supply several SP calculations.

A directed graph provides a compact mathematical representation of these
prerequisite relationships (\cref{fig:method-components-dependencies}). Each
box represents a calculation and each arrow points from a required earlier
result to the calculation that uses it. We refer to such a required earlier
calculation as a parent.
Snakemake \cite{Koster2012Snakemake,Molder2025SustainableSnakemake} uses this
graph to start independent branches when resources are available, wait for
their required inputs, and avoid repeating a compatible earlier calculation.

After resolving the YAML, EBSDmagus writes the planned calculations to a
single SQLite file \cite{SQLiteAbout}. The database associates each calculation
with its generated inputs and expected outputs. Before submission, EBSDmagus
compares the generated files, parent relationships, and planned job counts with
the YAML and database (\cref{sec:before-run-checks}). As jobs run, it adds their
execution history and checks whether each HDF5 output is readable and contains
the datasets required for its stage. It then compares the returned files and
completion records with the plan to establish which calculations finished and
which remain unresolved (\cref{sec:hpc-execution,sec:recovery-runs}). SQLite
requires no external database service, so the file remains with the run when it
is archived (\cref{sec:archiving}).

We submitted the calculations reported here through SLURM
\cite{Yoo2003SLURM}. To keep cluster choices out of the scientific description,
resource requests and other scheduler-specific settings are held in a separate
profile. Supporting another scheduler would mean providing an equivalent
profile and status mapping; the YAML and database would not otherwise change
(\cref{app:slurm-profiles}). \Cref{fig:method-components} summarises this
sequence and the branching between related calculations.

\Needspace{5\baselineskip}
During software development, we used OpenAI Codex to assist with code
generation and with developing and running unit, integration and system tests.
The authors defined the requirements, reviewed the resulting changes and test
outcomes, and retained responsibility for software design and scientific
validation.

\subsection{Preparing the Material Description}
\label{sec:material-description}
Every EMsoft calculation starts from a crystallographic structure. Depending on
its source, the structure may be available as a Crystallographic Information
File (CIF) \cite{Hall1991CIF} from a database or experiment, or as a POSCAR or
CONTCAR file from an atomistic calculation. EMsoft uses its own \texttt{.xtal}
format for the simulations considered here. EBSDmagus converts these source
formats into \texttt{.xtal} files. During CIF conversion, it rejects inputs from
which no atomic sites can be read and element symbols that cannot be resolved.
It also rejects space-group numbers outside the range 1--230 when assigning
the EMsoft crystal-system code. These checks do not independently establish
whether the space-group assignment and atomic arrangement describe the
intended material.

Separate conversion regressions compare lattice
parameters, space-group numbers, atomic sites, and occupancies with retained
reference data; \cref{sec:structure-input-tests,app:cif2xtal-fixtures} report
the comparisons and their present crystallographic coverage.

Temperature requires particular care because it is not entered as a direct
setting in the later EMsoft namelists. Its effect on thermal scattering is
represented through Debye--Waller values stored with the crystal data. For each
atomic site, EBSDmagus gives an explicit override precedence over an isotropic
displacement value supplied in the CIF. Where neither is supplied, it
calculates the value at the requested temperature from the Gao and Peng
parameterisation \cite{Gao1999DebyeWaller}. It converts CIF $B$ or $U$ values
to the units and convention required by EMsoft and rejects sites that specify
both, even when an override is provided. Changing the requested temperature
therefore does not replace supplied
CIF values; the researcher must choose an override if those values are not
appropriate for the intended study. These selection and conversion rules do
not assess whether a displacement value is physically suitable for the
material and temperature. EBSDmagus records the source of each selected value
alongside the crystal data.

The resulting material description contains the structure, site occupancies,
and scattering information. A comparison of beam energy may use the same
material description throughout, whereas a study of composition or atomic
ordering requires a separate checked structure for each intended state.
EBSDmagus retains the orientation convention when generating angle files, so
that the relationship between simulated patterns and crystal orientations
remains explicit. Consistent conventions are important when comparing
simulated and experimental patterns, for example using kikuchipy
\cite{Anes2026Kikuchipy} or AstroEBSD \cite{Foden2019RefinedTemplateMatching}.

\subsection{Defining the Study in YAML}
\label{sec:yaml-definition}
The YAML file brings the settings for the complete investigation into one
editable place. Without this common description, values that belong together
must be kept consistent across many stage-specific namelists and the separate
structure, Bethe-parameter, and Euler-angle files to which they refer. Common
settings and swept values appear together in the YAML, allowing each intended
variation to be stated once rather than introduced by repeatedly patching
near-identical EMsoft files.

The YAML references the prepared \texttt{.xtal} files through
\texttt{xtal\_files}; the CIF, POSCAR, or CONTCAR conversion precedes this step
(\cref{sec:material-description}). It adds the beam, diffraction, orientation,
and detector settings, including the Bethe parameters used to generate the
corresponding auxiliary files. Some of these values
govern which physical contributions enter the calculation. For example,
$d_{\min}$ is the smallest lattice-plane spacing included in the master-pattern
calculation and therefore changes the set of reflections considered. The
literal YAML key is \texttt{dmin}. The Bethe parameters provide numerical
cut-offs for classifying strong, weak, and double-diffraction beams, while the
beam energy determines the electron transport calculation.

For Bethe parameters, EBSDmagus checks the configuration format and the selected
mode for combining fixed and swept values. It generates a parameter file for
each requested combination and links it to the corresponding MP calculation;
the before-run checks confirm that the required files are present. It does not
assess whether the chosen cut-offs give converged diffraction results. The
researcher must establish suitable values for the intended calculation.

Other entries define
the requested Euler orientations and detector geometry. Fixed values and lists
to be swept use the same notation.

Because these parameters enter at different stages, changing them affects
different parts of the EMsoft sequence. A change of structure or beam energy
requires a new MC result and corresponding downstream calculations. Changing
$d_{\min}$ or the Bethe parameters requires a new MP calculation but may retain
a compatible MC result, whereas a different
orientation list or detector geometry requires only another SP calculation.
EBSDmagus applies these distinctions when it expands the YAML and selects the
matching earlier result for every generated call.

Each EMsoft call still receives its own NML and the files named by that NML.
EBSDmagus generates and checks this complete set, including the Euler-angle
files for SP calculations and the Bethe-parameter files for MP calculations.
Researchers review one YAML and a representative selection of the generated
inputs instead of manually aligning a large collection of nearly identical
files.
\Cref{fig:campaign-yaml-outline} shows a compact YAML example, while
\cref{tab:yaml-expansion} lists the files and records generated from it.

\begin{figure}[H]
  \centering
\begin{lstlisting}[language=yaml,basicstyle=\ttfamily\scriptsize]
# Same YAML file: fixed values and sweeps live together.
project_name: YNi2_2B_dmin_sweep
xtal_files: [YNi2.xtal]

monte_carlo:
  fixed:
    ekev: 30.0
    totnum_el: 2000000000

masterpattern:
  fixed:
    nthreads: 48
  sweep:
    dmin: [0.02, 0.05]

screen_pattern:
  fixed:
    scalingmode: gam
    gammavalue: 0.33
  anglefile_generator:
    values:
      - [0.0, 20.0, 0.0]
      - [325.85, 89.60, 118.63]
\end{lstlisting}
  \caption{Compact YAML excerpt. One source file records fixed values and swept parameters; EBSDmagus expands it into concrete EMsoft inputs, calculation graph, database entries, and before-run checks. \Cref{app:technical,fig:yaml-generated-files,fig:yaml-generated-checks} show fuller YAML excerpts and the generated files.}
  \label{fig:campaign-yaml-outline}
\end{figure}

\begin{table}[H]
  \centering
  \caption{Files and records generated from the YAML before submission and their relationships to the planned calculations.}
  \label{tab:yaml-expansion}
  \begin{tabular}{@{}>{\raggedright\arraybackslash}p{0.30\linewidth}>{\raggedright\arraybackslash}p{0.58\linewidth}@{}}
    \toprule
    Generated file or record & Information carried into the run \\
    \midrule
    MC, MP, and SP namelists &
    One NML file for each concrete calculation. The MC file names the checked \texttt{.xtal} input and its output; each MP file names its MC energy file and Bethe-parameter file; each SP file names the combined MP output containing the copied MC data, the angle file, and the final output. \\
    Angle files and SP settings &
    Orientation lists for the requested Euler angles or quaternions. The corresponding SP namelist records the angle convention, detector distance and tilt, pixel size and count, pattern centre, energy window, and intensity scaling. \\
    Bethe-parameter namelists &
    One file for each selected \texttt{c1}, \texttt{c2}, \texttt{c3}, and \texttt{sgdbdiff} combination, which controls the strong-, weak-, and double-diffraction beam selection. Each MP namelist points to the exact generated parameter file. \\
    Structure references and metadata &
    Links the generated calculations to the checked \texttt{.xtal} structure, resolved parameter values, conventions, and input/output paths needed for later checks and output-to-input traces. \\
    Database records &
    One record per unique MC, MP, or SP calculation gives its stage, project, resolved parameters, generated namelist, status, and hashes. Related records hold the expected output paths, check results, scheduler details, and status history. \\
    Calculation graph and before-run report &
    The graph shows every planned job, the MC parent of each MP calculation, and the combined MP output used by each SP calculation. The report compares the expected jobs and generated inputs with database records and output paths, and identifies missing, duplicate, or inconsistent records before submission. \\
    \bottomrule
  \end{tabular}
\end{table}

\subsection{Generating EMsoft Inputs and Run Records}
EMsoft expects each executable to receive a namelist in its own syntax together
with the stage-specific files named there. EBSDmagus generates these files from
Jinja \cite{PalletsJinjaDocs} templates for the MC, MP, and SP inputs instead of
relying on one-off substitutions. The templates preserve the exact layout
that EMsoft reads while keeping it separate from the compact YAML description.
If an EMsoft input format changes, the corresponding template can be updated
once instead of patching every prepared calculation. A different simulation
route can likewise be introduced through another set of templates and rules
without replacing the run database or the checking and reporting functions.

For each planned calculation, EBSDmagus derives a job hash from the stage,
crystal name, resolved parameters, and, for MP and SP, the exact parent hashes.
Overlapping branches within the same prepared study therefore use one job when
their definitions agree. Cluster settings such as partition and wall time are
recorded separately and do not alter this identity. A checksum detects later
changes to the generated NML, while the required angle and Bethe-parameter files
are checked for presence.

\subsection{Calculation Relationships and Result Reuse}
\label{sec:calculation-relationships}
The MC~$\rightarrow$~MP~$\rightarrow$~SP sequence reflects both the physical
separation in EMsoft and the order in which its executables run. The MC
calculation describes electron transport and its energy and depth
distributions. The MP calculation combines the matching MC result with the
crystal and scattering settings to calculate the master pattern, and writes a
combined HDF5 file containing the MC data needed during detector projection.
The SP calculation reads that combined file for both the transport and
master-pattern information. Because the combined MP file carries the MC data,
EMsoft executes a simple chain. The provenance record nevertheless retains
MC~$\rightarrow$~MP and (MC, MP)~$\rightarrow$~SP. If an SP output contains
several orientations, the database links the output and its ordered
orientation list to the same parent calculations.

\begin{figure}[H]
  \centering
  \includegraphics[width=0.98\linewidth]{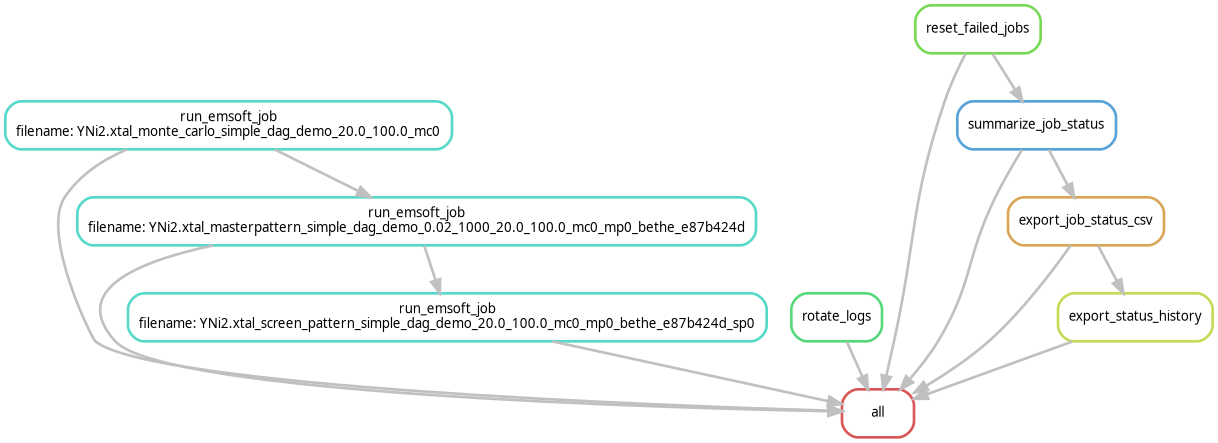}
  \caption{Calculation graph exported for a simple EBSDmagus run. The left-hand branch
  contains the EMsoft calculations: the MC output feeds the MP
  calculation, which produces the combined HDF5 input used by the SP stage.
  The right-hand branch contains supporting run-management tasks for resetting
  failed jobs, summarising and exporting status information, preserving status
  history, and rotating logs. \Cref{fig:large-preflight-dag-appendix}
  shows the larger parameter study described in \cref{sec:large-preflight}.}
  \label{fig:simple-dag}
\end{figure}

An MC output can feed several MP jobs only when its transport-defining material,
beam, and depth settings match. An MP output can feed several SP jobs when its
MP-defining inputs and exact MC parent match. By carrying the parent hashes into
the downstream job definitions,
EBSDmagus rejects an MP or SP calculation paired with a different upstream
result. EBSDmagus skips an exact repeated SP request, but a new orientation
list or detector setting
requires another screen-pattern calculation. \Cref{tab:stage-reuse} gives
the reuse conditions for the parameters introduced in
\cref{sec:yaml-definition}; the job hash uses the complete resolved parameter
set rather than only the examples listed in the table. The compact diagram in
\cref{fig:method-components-dependencies} explains the branching, while
\cref{fig:simple-dag} shows the graph exported from an actual simple run. The
exported version also contains the supporting tasks used to maintain the run
record.

\begin{table}[H]
  \centering
  \caption{Conditions for using an existing result at each EMsoft stage. The
  relevant physical and numerical settings are introduced in
  \cref{sec:yaml-definition}; the implemented identity uses the complete
  resolved parameter set and exact parent hashes.}
  \label{tab:stage-reuse}
  \begin{tabular}{@{}>{\raggedright\arraybackslash}p{0.09\linewidth}>{\raggedright\arraybackslash}p{0.52\linewidth}>{\raggedright\arraybackslash}p{0.30\linewidth}@{}}
    \toprule
    Stage & An existing result can be used when & A new result is required when \\
    \midrule
    MC &
    The checked material description and all transport-defining settings match. &
    The material description or any transport-defining setting changes. \\
    MP &
    The exact MC parent and all master-pattern settings match. The combined MP output carries the MC data needed for SP projection. &
    The MC parent or any master-pattern setting changes. \\
    SP &
    The exact combined MP parent and all projection settings match; an exact duplicate request can be skipped. &
    The parent, requested orientations, detector definition, or image settings change. \\
    \bottomrule
  \end{tabular}
\end{table}

\subsection{Checking Before Submission}
\label{sec:before-run-checks}
HPC projects receive a finite allocation, and every submitted job must also
pass through a shared queue. An incorrect calculation consumes time that cannot
be used for the intended study and may be discovered only after the job has
waited and run. EBSDmagus therefore checks the prepared work before requesting
cluster resources, while errors can still be corrected without repeating a
costly calculation.

The before-run check asks whether the files prepared by EBSDmagus represent the
study described in the YAML. It compares the expected number of MC, MP, and SP
jobs with the database, verifies their parent relationships and output paths,
recomputes the NML checksums, and confirms that the required inputs are present.
Missing, duplicate, or inconsistent records appear in an HTML report whose
opening summary states whether the work is ready for submission. The detailed
comparisons follow, together with the software and environment information
needed to reproduce the preparation and interpret the run.

\begin{figure}[H]
  \centering
  \includegraphics[width=0.94\linewidth]{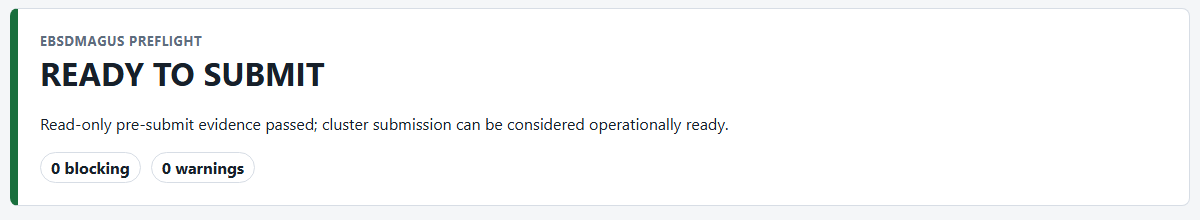}\\[-0.2em]
  {\small\textbf{(a)} Prepared run before modification}\par\medskip
  \includegraphics[width=0.94\linewidth]{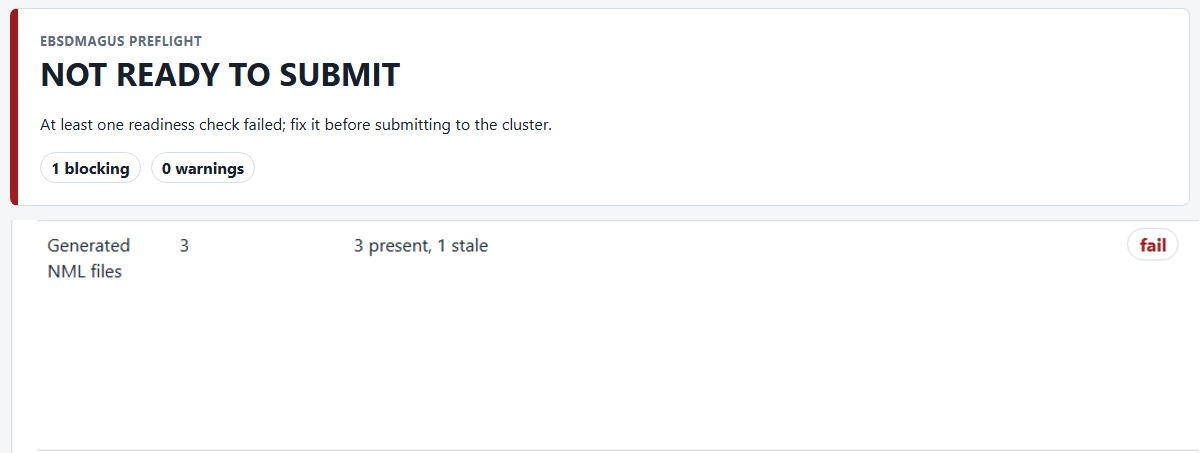}\\[-0.2em]
  {\small\textbf{(b)} Modified MC namelist detected before submission}
  \caption{Before-run reports generated for the same prepared run. (a) The
  original files passed all checks. (b) We then changed the incident beam
  energy in the generated MC namelist from \SI{20}{keV} to \SI{25}{keV}. The
  repeated check found one modified namelist and blocked submission; the
  other checks still passed. The complete HTML and JSON reports are included
  with the manuscript assets.}
  \label{fig:preflight}
\end{figure}

At this point, the graph describes the planned calculation order. The after-run
graph later marks each job as completed, incomplete, or failed.
\Cref{fig:preflight} shows the readiness summary used for the selected completed
run. A blocking mismatch prevents submission; \cref{tab:rejection-checks}
lists examples involving incompatible parents and incomplete HDF5 outputs.

\subsection{Running and Recovering on HPC}
\label{sec:hpc-execution}
Running the prepared calculations on a cluster differs from running them on a
desktop. The researcher requests shared compute nodes through a batch scheduler
and has less direct control over when and where a calculation starts. We tested
EBSDmagus with SLURM \cite{Yoo2003SLURM}, which is widely used on HPC systems.
The supplied cluster profile keeps queue names, partitions, wall times, and
module-loading details separate from the material and simulation settings.
Snakemake submits the calculations in the required order, while EBSDmagus
records the scheduler identifiers and output checks in the database.

Long-running workflow controllers cannot be assumed to remain available on
shared login nodes for the full duration of a study. Local usage limits,
scheduled maintenance, administrative cleanup, or failure of the controlling
process may stop them while jobs already submitted to SLURM continue on compute
nodes \cite{RWTHLoginNodesDocs,NERSCNextflowDocs}. Some computing centres
provide dedicated workflow queues or persistent
service hosts, but these facilities are site-specific rather than part of a
common HPC environment \cite{NERSCWorkflowQOSDocs}. Worker jobs can also fail or be cancelled while
independent branches remain runnable, and the shared filesystem can lag or fail
even when the scheduler reports success, leaving an HDF5 output delayed,
missing, or truncated. The cluster profile launches each EMsoft call through a
small EBSDmagus worker command. Besides starting the calculation, this command
records its exit code, expected output, file size, HDF5 structural checks,
compute node, time, and repository commit in a small completion file and a
machine-readable JSON record. EBSDmagus compares these records with the
database, scheduler history,
logs, and HDF5 outputs instead of relying on any one source.

The supplied cluster profile allows Snakemake to retry a failed worker job up to
two times and to continue work in independent branches. When database,
scheduler, and filesystem records disagree, EBSDmagus identifies calculations
with HDF5 files that passed
the defined structural checks, those that remain pending, and those that must
run again. The recovery commands then retain checked upstream results and reset
only the failed or incomplete branch. A job stopped by its wall-time or memory limit
remains failed; the researcher can adjust the cluster profile before resetting
and resubmitting it. EBSDmagus does not accept repeated failures, incompatible
parents, or incomplete files as completed work; it reports them for inspection.
\Cref{sec:recovery-runs} shows how EBSDmagus applied these checks after the
observed controller and shared-filesystem interruptions; the controlled worker-job
tests are documented in \cref{app:worker-cancellation}.

\subsection{Archiving and RDM Integration}
\label{sec:archiving}
Not every investigation proceeds to screen-pattern generation. Some require
only an MC result, others end with one or more master patterns, and studies of
orientation or detector settings also generate screen patterns. EBSDmagus
retains the requested stages and links each output to the inputs and earlier
calculations required to produce it. For each SP output, this includes the ordered
orientation list and its MC and MP parents.

Researchers can archive the database with these outputs. Together with the YAML,
generated reports, selected output checks, and recorded repository version, it
forms the portable package described in \cref{app:run-package}. A repository or
institutional RDM system can then add the persistent identifier, access
conditions, licence, and repository-level metadata. Other simulation packages
may divide their calculations differently; supporting one requires suitable
templates, dependency rules, and output checks rather than the specific
three-stage EMsoft arrangement.

\subsection{Using EBSDmagus}
The command-line interface and terminal user interface (TUI) call the same
preparation, checking, submission, monitoring, and recovery functions. Because
it runs in a terminal, users can access the TUI over SSH on a shared cluster and
follow these steps without memorising every command or needing a graphical
desktop. In either interface, the researcher selects the YAML and database,
reviews the before-run report, confirms submission, and inspects the after-run
result; EBSDmagus performs the file generation, comparisons, and status updates
requested at each step.
\Cref{app:user-route,app:tui-screenshots} provide the practical checklist and
example screenshots.

\section{Configuration, Execution, and Recovery Tests}
\label{sec:validation-runs}

We tested whether EBSDmagus preserves structure data during conversion,
prepares and completes the requested calculations, and retains completed work
when recovering interrupted runs. The cases comprise structure-conversion
regressions, a complete
MC~$\rightarrow$~MP~$\rightarrow$~SP calculation for YNi$_2$, preparation of a
530-job parameter study, and recovery from interrupted cluster runs. The larger
example was checked without consuming the corresponding cluster allocation;
the recovery tests include two interruptions observed during normal runs and
controlled worker-job cancellations. Complete configurations and run records
are given in \cref{app:technical,app:run-details}.

\subsection{Structure-Input Regression Tests}
\label{sec:structure-input-tests}
We regenerated 16 retained CIF-to-\texttt{.xtal} reference sets spanning eight
space-group numbers in four crystal systems. For each set, the atom data,
lattice parameters, space-group number, and atom types agreed with the recorded
HDF5 reference values. We also converted ten retained Mg Laves-phase CONTCAR
files associated with the defect studies of Tehranchi et al.
\cite{Tehranchi2023LavesDefects} to CIF and recovered the atom count, species
order, cell parameters, and
wrapped fractional coordinates in the paired CIF references. Because these
pairs have no independent \texttt{.xtal} references, they test the
CONTCAR-to-CIF conversion only. The retained reference sets provide repeatable
regression coverage for the cases listed in
\cref{app:cif2xtal-fixtures}, but do not cover all crystallographic settings.

\subsection{Completed Representative Run}
\label{sec:completed-run}
\begin{figure}[htbp]
  \centering
  \begin{subfigure}[t]{0.28\linewidth}
    \centering
    \includegraphics[width=\linewidth,height=3.45cm,keepaspectratio]{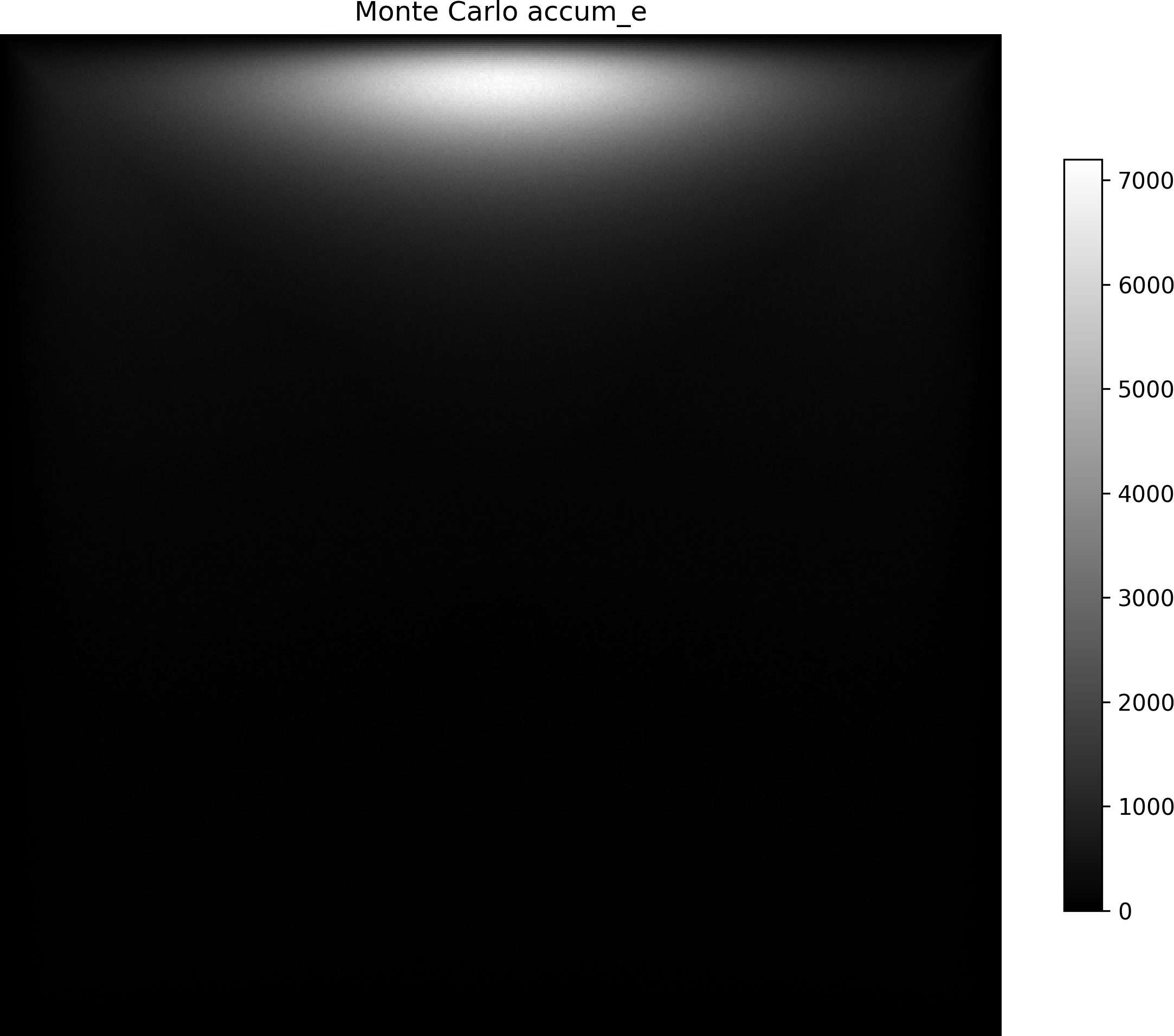}\\[-0.1em]
    \textbf{(a)}
  \end{subfigure}
  \hfill
  \begin{subfigure}[t]{0.28\linewidth}
    \centering
    \includegraphics[width=\linewidth,height=3.45cm,keepaspectratio]{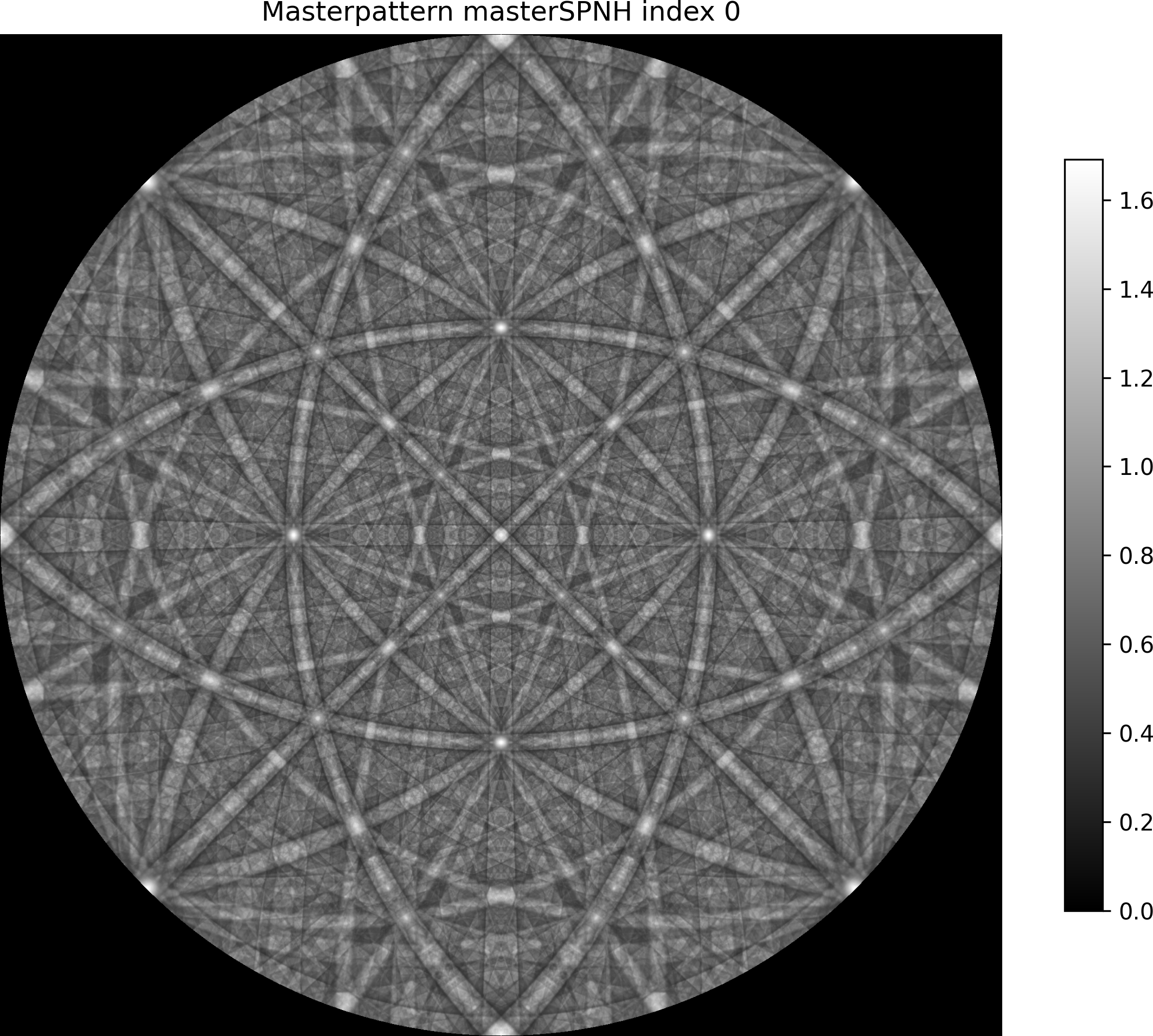}\\[-0.1em]
    \textbf{(b)}
  \end{subfigure}
  \hfill
  \begin{subfigure}[t]{0.39\linewidth}
    \centering
    \includegraphics[width=\linewidth,height=3.45cm,keepaspectratio]{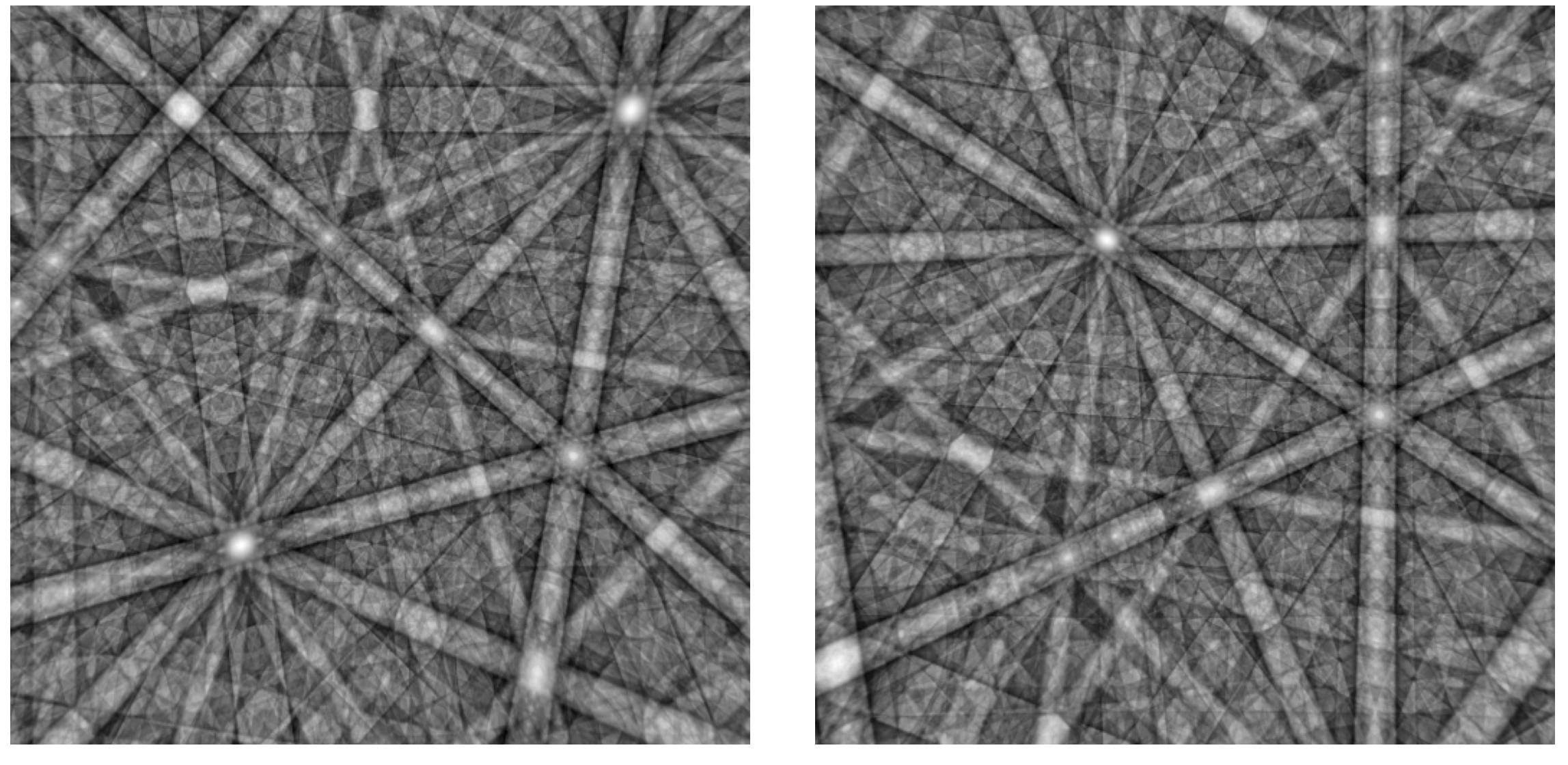}\\[-0.1em]
    \textbf{(c)}
  \end{subfigure}
  \caption{Selected outputs from the representative completed run:
  (a) accumulated Monte Carlo signal, (b) master-pattern preview, and
  (c) two screen patterns.}
  \label{fig:output-previews}
\end{figure}

The representative run contains one MC calculation, two MP calculations, and
two SP calculations. The two MP branches use $d_{\min}$ values of
\SIlist{0.02;0.05}{nm}, where $d_{\min}$ is the smallest lattice-plane spacing
included in the master-pattern calculation. The run used the complete
EMsoft execution path, with a restricted \SIrange{29.5}{30}{keV} energy window
to limit its computational cost. All five calculations completed, and their
HDF5 outputs passed the defined structural checks. Neither the before-run nor
the after-run report contained blocking findings or warnings. This run covers
the complete prepared calculation sequence and its retained records. Assessing
the physical effect of $d_{\min}$ in YNi$_2$ would require a broader parameter
study.
\Cref{fig:output-previews} shows representative MC, MP, and SP outputs from this
run.

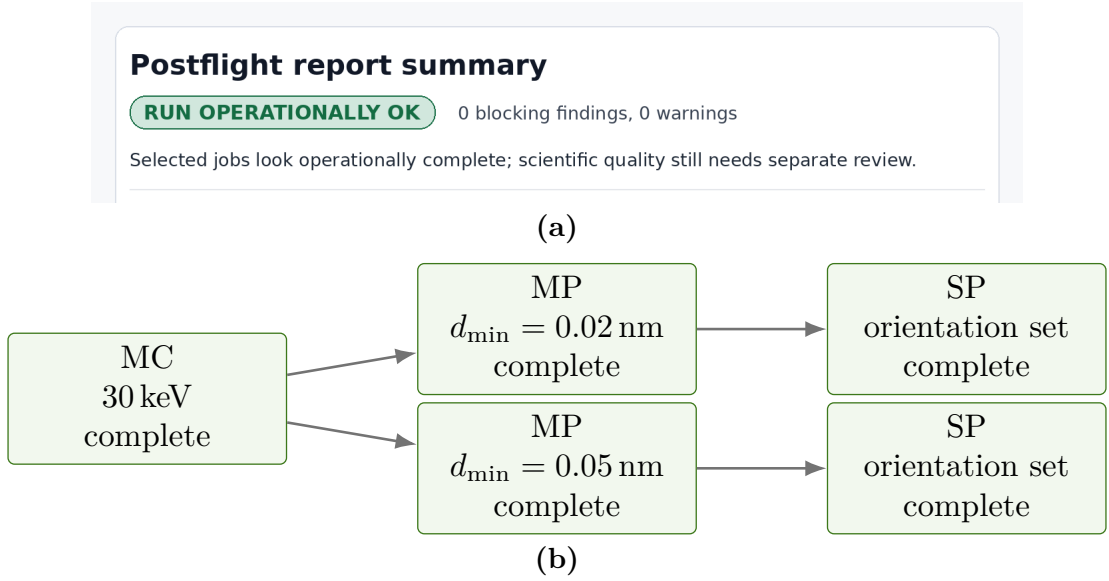
\begin{figure}[H]
  \centering
  \includegraphics[width=0.78\linewidth,trim=0 350bp 0 0,clip]{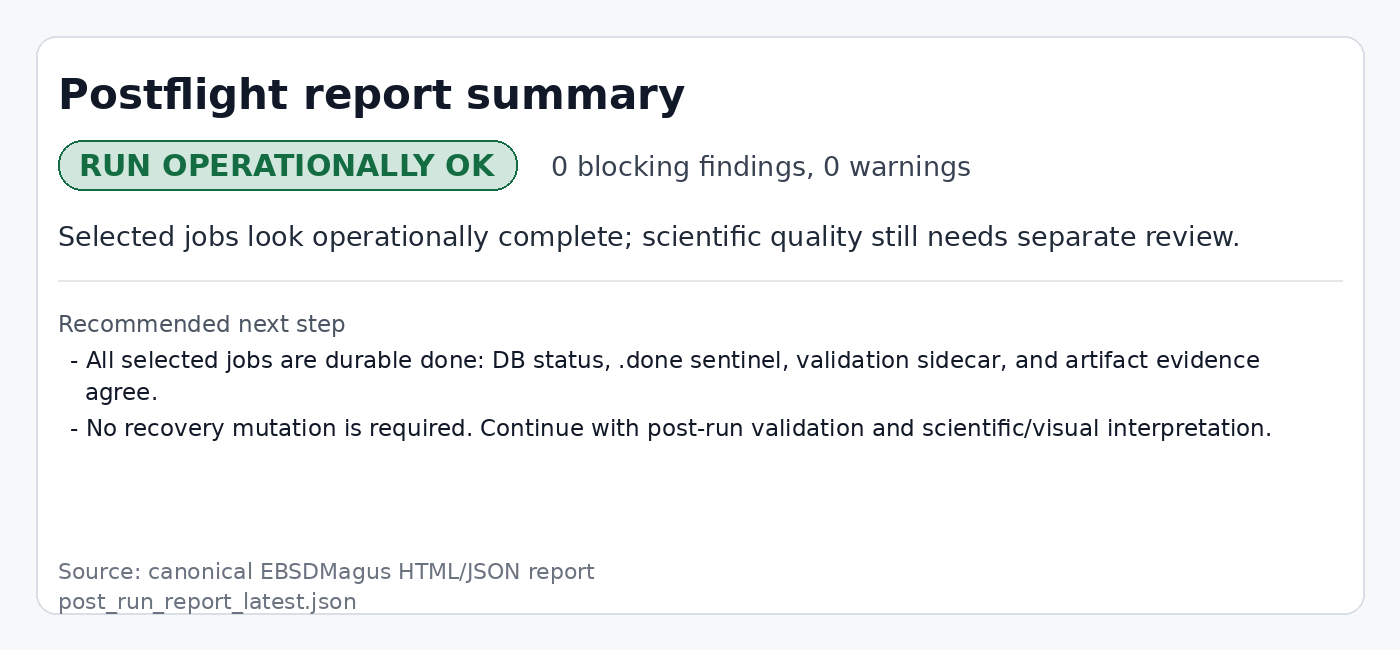}\\[-0.1em]
  \textbf{(a)}

  \vspace{0.6em}
  \resizebox{0.92\linewidth}{!}{%
  \begin{tikzpicture}[
    done/.style={draw=ebsdgreen!70!black,fill=ebsdgreen!8,rounded corners=2pt,
      minimum width=3.0cm,minimum height=0.85cm,align=center,font=\small},
    statusArrow/.style={-Latex,thick,draw=black!55}
  ]
    \node[done] (mc) {MC\\\SI{30}{keV}\\complete};
    \node[done,right=1.4cm of mc,yshift=0.75cm] (mpa)
      {MP\\$d_{\min}=\SI{0.02}{nm}$\\complete};
    \node[done,right=1.4cm of mc,yshift=-0.75cm] (mpb)
      {MP\\$d_{\min}=\SI{0.05}{nm}$\\complete};
    \node[done,right=1.4cm of mpa] (spa) {SP\\orientation set\\complete};
    \node[done,right=1.4cm of mpb] (spb) {SP\\orientation set\\complete};
    \draw[statusArrow] (mc) -- (mpa);
    \draw[statusArrow] (mc) -- (mpb);
    \draw[statusArrow] (mpa) -- (spa);
    \draw[statusArrow] (mpb) -- (spb);
  \end{tikzpicture}}\\[-0.1em]
  \textbf{(b)}
  \caption{After-run check for the representative calculation. (a) The report
  contains no blocking findings or warnings. (b) All five EMsoft calculations
  completed (green); their HDF5 outputs passed the structural checks. We shortened the
  filenames and omitted status and log tasks in panel (b); the full graph is
  archived.}
  \label{fig:postflight}
\end{figure}

After execution, EBSDmagus found no discrepancies among the database, scheduler
history, worker records, and HDF5 outputs. Optional
previews support visual inspection, but the completion decision uses the
recorded execution result and structural HDF5 checks. The calculation graph
then shows which EMsoft jobs completed, remain pending, or need attention.
The corresponding summary and calculation-status view appear in
\cref{fig:postflight}.

For one SP output, we followed the recorded links through its MP and MC parents
to the run-local YAML. The trace includes the ordered orientation list,
generated NML, detector geometry, parent calculations, and the checks used to
accept the output. These records identify the input choices behind the SP file
before its patterns are interpreted. The database stores these links in the
tables shown in
\cref{fig:database-schema-overview,tab:dba-schema-reading-guide,tab:rdm-map}.
In the example used here, the final SP output links to its generated NML and
angle settings, the MP calculation with \texttt{dmin=0.02},
\texttt{npx=1000}, and the selected Bethe parameters, and the parent MC
calculation at \SI{30}{keV}. The archived trace retains the full filenames,
hashes, and paths as part of the run package described in
\cref{app:run-package}.

\subsection{Preparation of a Larger Example Study}
\label{sec:large-preflight}
The larger example varies the beam energy between \SIlist{5;10;15;20;30}{keV}
and uses two electron-transport depth limits at each energy. These ten MC
calculations feed 40 MP calculations obtained from two $d_{\min}$ values
(\SIlist{0.02;0.05}{nm}) and two Bethe-parameter settings. Variations in the
pattern centre and gamma intensity scaling then produce 480 SP jobs. Each SP
job contains 200 Euler triplets, corresponding to 96,000 detector patterns if
the complete study were executed.

Before requesting cluster time, we generated the stage-specific EMsoft inputs
for this example and compared them with the
database and calculation graph. The before-run report found all 530
planned EMsoft jobs, with no duplicated output names, missing dependencies, or
orphaned downstream calculations. We did not submit the study because work of
this size requires a scientific allocation and an associated materials
question. \Cref{tab:large-preflight} summarises the expansion, while
\cref{fig:large-preflight-dag-appendix} shows how the parameter choices fan out
through the three calculation stages. The full graph, YAML file, and
before-run report are included with the supporting material.

\begin{table}[H]
  \centering
  \caption{Larger multi-parameter example prepared and checked before submission.
  \Cref{fig:large-preflight-dag-appendix} shows the branching from 10 MC to 40
  MP and 480 SP calculations; the supporting material contains the complete
  graph.}
  \label{tab:large-preflight}
  \begin{tabular}{@{}p{0.34\linewidth}p{0.56\linewidth}@{}}
    \toprule
    Report item & Current value \\
    \midrule
    Planned EMsoft jobs & 530 \\
    Unique Monte Carlo jobs & 10: five beam energies $\times$ two depth limits \\
    Unique master-pattern jobs & 40: four branches per MC calculation \\
    Screen-pattern jobs & 480: twelve detector and gamma-scaling branches per MP calculation \\
    Implied detector patterns & 480 SP jobs $\times$ 200 Euler triplets per angle file = 96,000 \\
    Most MP calculations sharing one MC output & 4 master-pattern branches \\
    Most SP calculations sharing one MP output & 12 screen-pattern branches \\
    Calculation graph & 536 nodes and 1,588 edges, including six preparation and checking steps \\
    Result of the before-run check & Ready for submission; no blocking findings or warnings \\
    Execution status & Prepared and checked; not submitted \\
    \bottomrule
  \end{tabular}
\end{table}
\subsection{Recovering Interrupted Cluster Runs}
\label{sec:recovery-runs}
On an HPC cluster, the calculation controller, worker nodes, and shared
filesystem are separate components. The controller can terminate while jobs
already submitted to SLURM continue, a worker can fail while independent
branches remain runnable, and the scheduler can report success even though an
output is delayed or lost on the filesystem. EBSDmagus therefore compares the
database, scheduler history, worker records, and HDF5 files before deciding
which calculations have completed.

\begin{figure}[H]
  \centering
  \resizebox{0.96\linewidth}{!}{%
  \begin{tikzpicture}[
    attention/.style={draw=red!70!black,fill=red!7,rounded corners=2pt,
      minimum width=3.3cm,minimum height=0.9cm,align=center,font=\scriptsize},
    waiting/.style={draw=black!45,fill=black!4,rounded corners=2pt,
      minimum width=2.8cm,minimum height=0.8cm,align=center,font=\scriptsize},
    recoveryArrow/.style={-Latex,thick,draw=black!50}
  ]
    \node[attention] (mc) {MC\\database: pending\\worker/output: complete};
    \node[waiting,right=1.15cm of mc,yshift=1.5cm] (mp1) {MP\\$d_{\min}=\SI{0.02}{nm}$, Bethe A};
    \node[waiting,right=1.15cm of mc,yshift=0.5cm] (mp2) {MP\\$d_{\min}=\SI{0.02}{nm}$, Bethe B};
    \node[waiting,right=1.15cm of mc,yshift=-0.5cm] (mp3) {MP\\$d_{\min}=\SI{0.05}{nm}$, Bethe A};
    \node[waiting,right=1.15cm of mc,yshift=-1.5cm] (mp4) {MP\\$d_{\min}=\SI{0.05}{nm}$, Bethe B};
    \node[waiting,right=1.15cm of mp1] (sp1) {SP A\\not submitted};
    \node[waiting,right=1.15cm of mp2] (sp2) {SP B\\not submitted};
    \node[waiting,right=1.15cm of mp3] (sp3) {SP C\\not submitted};
    \node[waiting,right=1.15cm of mp4] (sp4) {SP D\\not submitted};
    \foreach \m in {mp1,mp2,mp3,mp4} \draw[recoveryArrow] (mc) -- (\m);
    \draw[recoveryArrow] (mp1) -- (sp1);
    \draw[recoveryArrow] (mp2) -- (sp2);
    \draw[recoveryArrow] (mp3) -- (sp3);
    \draw[recoveryArrow] (mp4) -- (sp4);
  \end{tikzpicture}}\\[-0.1em]
  \textbf{(a)}

  \vspace{0.7em}
  \resizebox{0.96\linewidth}{!}{%
  \begin{tikzpicture}[
    recovered/.style={draw=ebsdgreen!70!black,fill=ebsdgreen!8,rounded corners=2pt,
      minimum width=2.8cm,minimum height=0.8cm,align=center,font=\scriptsize},
    recoveryArrow/.style={-Latex,thick,draw=black!50}
  ]
    \node[recovered,minimum width=3.3cm] (mc) {MC\\complete};
    \node[recovered,right=1.15cm of mc,yshift=2.1cm] (mp1) {MP\\$d_{\min}=\SI{0.02}{nm}$, Bethe A\\complete};
    \node[recovered,right=1.15cm of mc,yshift=0.7cm] (mp2) {MP\\$d_{\min}=\SI{0.02}{nm}$, Bethe B\\complete};
    \node[recovered,right=1.15cm of mc,yshift=-0.7cm] (mp3) {MP\\$d_{\min}=\SI{0.05}{nm}$, Bethe A\\complete};
    \node[recovered,right=1.15cm of mc,yshift=-2.1cm] (mp4) {MP\\$d_{\min}=\SI{0.05}{nm}$, Bethe B\\complete};
    \node[recovered,right=1.15cm of mp1] (sp1) {SP A\\complete};
    \node[recovered,right=1.15cm of mp2] (sp2) {SP B\\complete};
    \node[recovered,right=1.15cm of mp3] (sp3) {SP C\\complete};
    \node[recovered,right=1.15cm of mp4] (sp4) {SP D\\complete};
    \foreach \m in {mp1,mp2,mp3,mp4} \draw[recoveryArrow] (mc) -- (\m);
    \draw[recoveryArrow] (mp1) -- (sp1);
    \draw[recoveryArrow] (mp2) -- (sp2);
    \draw[recoveryArrow] (mp3) -- (sp3);
    \draw[recoveryArrow] (mp4) -- (sp4);
  \end{tikzpicture}}\\[-0.1em]
  \textbf{(b)}
  \caption{Observed controller interruption and recovery. (a) The database
  marked MC as pending although the worker record and HDF5 output showed
  completion; red marks this disagreement. Four MP and four SP calculations had
  not yet run (grey). (b) The recovery
  command updated the MC record and resumed the remaining calculations; all nine
  completed (green). Full filenames and intermediate reports are archived.}
  \label{fig:controller-loss-recovery}
\end{figure}
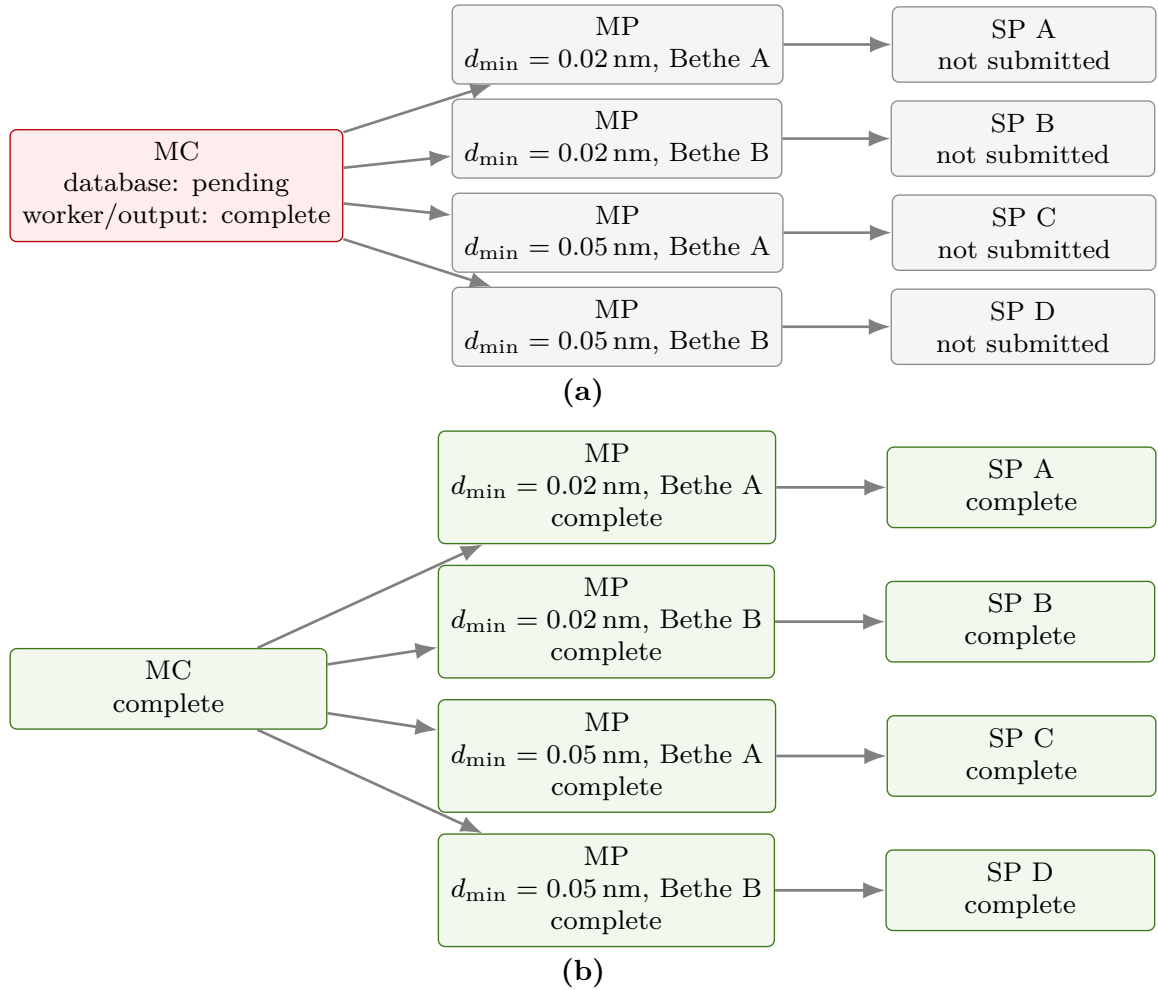

We first observed controller loss when the MC worker completed but the
Snakemake process was no longer running before it updated the database or
submitted the later MP and SP calculations. The database still listed MC as pending,
whereas SLURM, the worker records, and the structurally readable HDF5 output
agreed that it had completed. The preserved records did not establish why the
controller ended, and we found no indication of a login-node reboot. The
documented recovery command updated the stale MC database entry to agree with
the checked worker result, retained its output, and resumed the remaining
calculations without repeating MC. All nine EMsoft jobs completed after
recovery; \cref{fig:controller-loss-recovery} shows the initial
disagreement and the final result.

A later full-chain test exposed a separate shared-filesystem failure. Snakemake
reported repeated ``stale file handle'' errors while accessing its logs and run
files, and the controller was no longer running shortly afterwards. An MP job
already executing on a worker completed, but the database still listed it as pending
and SP had not been submitted. We preserved the run before recovery. The
scheduler and worker records agreed with the MC and MP HDF5 files, both of which
passed the structural checks, so EBSDmagus retained them and submitted only the
missing SP calculation. The final after-run check found all three stages
complete. The logs establish that the filesystem became unavailable to the
controller, but not whether it caused the preceding scheduler error. The two
observed interruptions and the files retained for their analysis are separated
in \cref{app:observed-interruptions}.

A worker failure produces a different pattern. The controller can either stop
scheduling new work after the first error or continue with independent branches;
dependent calculations remain blocked in both cases. EBSDmagus uses the latter
behaviour by default. In controlled tests, we cancelled one running MP job. The
independent branch completed, while the cancelled MP and its dependent SP
remained unresolved until the documented reset and recovery commands resumed
them. The alternative stop-on-error behaviour and both recovery sequences are
reported in \cref{app:worker-cancellation}.

EBSDmagus accepts an output for further use only when its
parent identity is compatible and the database, worker result, and expected
HDF5 file agree after comparison. The HDF5 check confirms that the file is non-empty,
readable, and contains the dataset required for its MC, MP, or SP stage;
conflicting or incomplete records remain marked for attention. The researcher
then inspects the physical content separately. \Cref{tab:rejection-checks}
lists the parent and output mismatches rejected by the current checks, while
\cref{tab:database-status-vocabulary,fig:database-schema-overview} describe
the corresponding database records.

\section{Benefits and Limitations of EBSDmagus}
\label{sec:discussion}

An individual EMsoft calculation takes the same time with or without
EBSDmagus; the practical saving lies in preparing and following many related
calculations through several queueing cycles and recovering them after an
interruption. Without a common definition and run record, researchers must
repeatedly compare scheduler entries, logs, and output directories to decide
which result belongs to which input and what still needs to run.

In the larger example, five beam energies and two transport depths produce ten
MC calculations. Two values of $d_{\min}$ and two Bethe settings expand these
to 40 MP calculations, followed by 480 SP jobs and 96,000 detector patterns.
The researcher specifies these variations in one YAML rather than assembling
and cross-checking 530 configurations separately. The campaign-level benefit
extends beyond generating many orientation projections: the settings remain
linked to compatible MC and MP results, execution records, and the decisions
used to retain or repeat calculations after an interruption.

When several material and experimental variables are considered together, the
scientific choices may still fit into a short parameter list, while the number
of NML files, auxiliary inputs, and dependent calculations grows much faster.
EBSDmagus expands the choices into compatible MC, MP, and SP branches and checks
their relationships before cluster time is requested. A suitable compute
allocation remains necessary, but file preparation and parent selection no
longer have to be repeated by hand for every branch.

A companion Au--Ni study used EBSDmagus for simulations that varied
composition, lattice parameter, atomic ordering, and diffraction settings, and
compared selected patterns with experimental EBSD data
\cite{Teixeira2026AuNiKikuchi}.

The time saved cannot be reduced to one generally applicable number: it depends
on the size of the study, queue delays, failure frequency, and local experience.
In the interrupted runs, EBSDmagus retained the completed MC result after the
first controller termination, retained MC and MP after the shared-filesystem
interruption, and allowed independent branches to finish during the controlled
worker cancellation. Recovery therefore required resubmitting unresolved work,
rather than continuously monitoring every job or repeating upstream
calculations whose outputs had already passed the checks.

The records used during recovery also document how an archived output was
produced. They link it to the resolved settings, generated EMsoft inputs, and
upstream calculations, so publication does not depend on reconstructing this
history from filenames and personal notes. The YAML, SQLite, JSON, and HDF5
files can accompany the simulated patterns into a repository or institutional
RDM system, where persistent identifiers, access conditions, licences, and
repository-level metadata can be added
\cite{Wilkinson2016FAIR,NFDIMatWerkProject}. The database and package map are
given in \cref{app:database,app:run-package}.

The present tests cover the three-stage EMsoft route on one SLURM-managed HPC
system. Other simulators or schedulers would require their own templates,
calculation rules, output checks, or cluster profiles and would have to be
tested on the target system. Resource limits also remain study-specific: if a
job exceeds its wall-time or memory allocation, the cluster profile must be
adjusted before that calculation is resumed.

\section{Conclusion}
\label{sec:conclusion}

EBSDmagus keeps a dynamical EBSD parameter study coherent from its definition
through EMsoft execution and later archiving. The YAML description remains
connected to the concrete inputs, the MC, MP, and SP relationships, the cluster
history, and the resulting files, so these elements do not have to be
coordinated separately as the number of calculations grows.

The representative YNi$_2$ calculation completed the full three-stage route,
and the larger before-run example prepared 530 linked EMsoft jobs representing
96,000 detector patterns. The crystallographic regression tests reproduced the
retained reference quantities within their stated coverage. During two
observed cluster interruptions and a controlled worker cancellation, the
recorded checks distinguished completed work from unresolved branches and
allowed the latter to resume without discarding valid upstream results.

For the tested EMsoft and SLURM setup, the material and simulation choices
remain linked to the checked outputs and the archived record. Larger studies
therefore require less manual file coordination and job monitoring, while the
information needed to inspect or reuse the resulting patterns is retained.
\section*{Software Availability}
EBSDmagus is available under the MIT licence at
\url{https://gitlab.git.nrw/rwth-sfb1394/sfb1394/ebsd/ebsdmagus}.
The version prepared for this manuscript is v1.3. Its source archive is named
\path{ebsdmagus-v1.3-source.zip}. The software and supporting records are
prepared for deposition on Zenodo at
\url{https://zenodo.org/records/21646280}.

\section*{Data and Supporting Records Availability}
The computational inputs, selected outputs, run databases, check records, and
manifests used in this study are collected in
\path{ebsdmagus-v1.3-manuscript-supporting-records.zip}. This file will be
deposited in the same versioned Zenodo release as the software source. It
contains the study YAML files, database snapshots, before- and after-run
reports, calculation graphs, provenance records, test results, manifests and
checksums, and selected generated outputs and worker check records, as described
in \cref{app:run-package}.


\section*{Funding}
This work was supported by the German Research Foundation (DFG) within the Collaborative Research Centre SFB 1394 "Structural and Chemical Atomic Complexity—From Defect Phase Diagrams to Materials Properties" (Project ID 409476157), projects A07 and C02.  
This project has received funding from the European Research Council (ERC) under the European Union’s Horizon 2020 research and innovation programme (grant agreement No. 101168203 TailorPlast) (SKK). 

The authors gratefully acknowledge the computing time provided to them at the NHR Center NHR4CES at RWTH Aachen University (project number p0020330). This is funded by the Federal Ministry of Education and Research, and the state governments participating on the basis of the resolutions of the GWK for national high performance computing at universities (www.nhr-verein.de/unsere-partner).

The data used in this publication were managed using the framework and metadata
scheme provided by project A07 within SFB 1394 ``Structural and Chemical Atomic
Complexity -- From Defect Phase Diagrams to Material Properties'', project ID
409476157, funded by the Deutsche Forschungsgemeinschaft (DFG, German Research
Foundation).
The data were managed using the research data management platform Coscine with storage space granted by the Research Data Storage (RDS) of the DFG and Ministry of Culture and Science of the State of North Rhine-Westphalia (DFG: INST222/1261-1 and MKW: 214-4.06.05.08 - 139057). 

\section*{CRediT Author Statement}
Ulrich Kerzel: Conceptualization, Methodology, Software, Validation,
Investigation, Data curation, Visualization, Writing -- original draft,
Writing -- review \& editing, Project administration, Funding acquisition, Supervision. \\
Lukas Berners:
Methodology, Validation, Writing -- review \& editing. \\
Sandra Korte-Kerzel:
Writing -- review \& editing, Funding acquisition, Supervision.

\appendix
\counterwithin{figure}{section}
\counterwithin{table}{section}
\clearpage
\section{Practical User Checklist}
\label{app:user-route}

A new user normally begins with an example YAML and prepares the database before
generating and checking the EMsoft inputs. The same run record is used for
submission, monitoring, recovery, and archiving. Researchers select the
structure, physical model, and numerical settings using their EBSD knowledge
and, where necessary, the EMsoft documentation; \cref{tab:user-checklist}
gives the corresponding commands and checks.

\begin{longtable}{p{0.18\linewidth}p{0.34\linewidth}p{0.38\linewidth}}
\caption{Practical sequence for preparing, checking, submitting, recovering, and archiving an EBSDmagus run.}
\label{tab:user-checklist}\\
\toprule
\textbf{Step} & \textbf{What the researcher edits or checks} & \textbf{Typical command or output} \\
\midrule
Choose a starting point &
Select an example YAML file close to the intended material and study type. &
Example files under \texttt{workflow\_configs/}. \\
Prepare the database &
Create or select the SQLite database explicitly. &
\texttt{poetry run ebsdm init-db --database workflow.db}. \\
Describe the study &
Edit material/structure inputs, stage-specific MC, MP, and SP settings, fixed parameters, sweeps, orientations, detector settings, and the intended project name in the YAML. &
The YAML is the human-edited study description. \\
Generate EMsoft inputs &
Render concrete EMsoft NML files, angle files, Bethe-parameter files, metadata files, run-local profile/config files, and database rows. &
\texttt{poetry run ebsdm prepare --config workflow\_configs/<run>.yaml}. \\
Test locally where possible &
Check the generated Snakemake setup without spending cluster time; developers can also use Mock-SLURM (\cref{app:mock-slurm}) for local scheduler-submission tests. &
\texttt{poetry run ebsdm check --deep --html}; local Snakemake dry-runs and Mock-SLURM tests support development, while live-cluster validation remains necessary for real submissions. \\
Check before submission &
Inspect generated files, calculation graph, database rows, expected outputs, and the before-run report before using cluster time. &
\texttt{poetry run ebsdm check --preflight}; inspect the HTML report summary, calculation graph, and expected-vs-found tables. \\
Preview submission &
Check the exact command for the selected cluster profile. &
\texttt{poetry run ebsdm submit}. \\
Submit explicitly &
Submit only after the before-run checks and preview look correct. &
\texttt{poetry run ebsdm submit --confirm}. \\
Monitor execution &
Follow the database and scheduler records while Snakemake is active. &
Use the TUI/monitoring commands or scheduler views to check queued, running, failed, and completed jobs. \\
Recover if needed &
If jobs, scheduler records, database rows, or output files disagree, preserve the current run information and use the recovery commands before any manual database edits. &
Use the report recommendations first; expert recovery tools reconcile records, reset affected jobs, and resume the remaining calculations. \\
Verify jobs and outputs &
Check that the expected jobs have completion records and that dependent calculations have structurally readable HDF5 inputs/outputs. &
Use the after-run report to compare database status, scheduler records, \texttt{.done} markers, validation files, and HDF5 checks. \\
Inspect after the run &
Generate a bounded set of optional preview images, then regenerate the after-run report. The preview manifest records the exact MC, MP, and SP calculations and selected orientations represented by the images. &
\texttt{poetry run ebsdm post-run-check --html}; \texttt{poetry run ebsdm preview}; rerun \texttt{post-run-check --html}. \\
Archive or hand over &
Keep the YAML, database, before-/after-run reports, calculation graphs, provenance traces, source structure files, selected previews, and version/tag with the data package. &
Use this package for deposit in a data repository such as Zenodo or for institutional RDM integration. \\
\bottomrule
\end{longtable}

\clearpage
\section{Study Definition and Generated Inputs}
\label{app:technical}

EBSDmagus expands one study YAML into the files and records used by EMsoft and
Snakemake during execution and by the checking, recovery, and archiving
commands.

\subsection{Structure-Conversion Reference Checks}
\label{app:cif2xtal-fixtures}

The structure-input regressions in \cref{sec:structure-input-tests} compare
converted structures with retained reference files. The CIF conversion set
contains 16 complete CIF/\texttt{.xtal}/HDF5-dump
triplets. For every triplet, the same check converts the CIF to a temporary
\texttt{.xtal} file, compares its atom data with the recorded HDF5 values,
checks the supplied reference \texttt{.xtal} against the same record, and then
compares the extracted lattice parameters, space-group number, atom types, and
atom data. All 64 comparisons passed for the software version used in this
work. \Cref{tab:cif2xtal-fixtures} summarises the crystallographic range of the
set.

\begin{table}[htbp]
  \centering
  \caption{Crystallographic coverage of the retained CIF-to-\texttt{.xtal} reference sets. Each complete triplet passed the four conversion and metadata comparisons described in the text.}
  \label{tab:cif2xtal-fixtures}
  \begin{tabular}{@{}lcc@{}}
    \toprule
    Crystal system & Complete reference sets & Space-group numbers \\
    \midrule
    Orthorhombic & 3 & 59, 60 \\
    Tetragonal & 2 & 136, 139 \\
    Hexagonal & 7 & 187, 193, 194 \\
    Cubic & 4 & 227 \\
    \bottomrule
  \end{tabular}
\end{table}

The fixtures use EMsoft space-group setting 1 and cover the corresponding
standard settings. The current CIF converter assigns setting 1 without warning
the user about this assumption. Alternative origin choices and non-standard
axis settings remain untested. Before simulation, the researcher must verify
that the CIF coordinates use the origin and axes expected by EMsoft; successful
conversion alone does not establish this agreement.

A second regression covers the ten Mg Laves-phase CONTCAR/CIF pairs associated
with the work of Tehranchi et al. \cite{Tehranchi2023LavesDefects}. For
each pair, the converter writes a temporary CIF and compares it with the
retained reference. The comparison checks atom count, species order, all six
cell parameters, and wrapped fractional coordinates. All ten pairs passed for
the software version used in this work, and the test leaves the fixture
directory unchanged. Because these pairs do not include independently retained
\texttt{.xtal} files, they establish the CONTCAR-to-CIF comparison but not the
complete CONTCAR-to-\texttt{.xtal} conversion sequence.

\subsection{Simple Three-Stage YAML}
The simple example shows the MC~$\rightarrow$~MP~$\rightarrow$~SP sequence without parameter
branching. The full YAML remains in the archived run package; the excerpt below
shows how the file carries the relevant settings when no parameter is
swept. The bold orange \texttt{sweep} keys mark the empty sweep lists in this
single-valued example.

\begin{lstlisting}[
  language=yaml,
  emph={sweep},
  emphstyle=\bfseries\color{ebsdorange!85!black}
]
# Simple MC -> MP -> SP run: the essential source-of-truth fields.
project_name: simple_dag_demo
variation_mode: grid
xtal_files: [YNi2.xtal]
enabled_stages: [monte_carlo, masterpattern, screen_pattern]

monte_carlo:
  fixed:
    ekev: 20.0
    depthmax: 100.0
    totnum_el: 300000
  sweep: {}

masterpattern:
  fixed:
    dmin: 0.02
    npx: 1000
    nthreads: 48
  sweep: {}

screen_pattern:
  fixed:
    numsx: 514
    numsy: 514
    eulerconvention: tsl
    scalingmode: not
  anglefile_generator:
    mode: eu
    values:
      - [0.0, 20.0, 0.0]
      - [325.85, 89.60, 118.63]
  sweep: {}
\end{lstlisting}

\yamllistingneed
\subsection{Selected \texorpdfstring{\SI{2e9}{}}{2e9}-Electron Sweep Excerpt}
The selected executed run uses the YNi2 \SI{2e9}{} electron configuration with
two representative \texttt{dmin} settings and two Euler angles. The excerpt shows
only the differences that matter for this run: electron count, master-pattern
\texttt{dmin}, detector scaling, and the two orientations. The bold orange
\texttt{dmin} entry is the YAML change that creates two MP branches from the
same MC output. The full YAML source remains in the archived run package.

\begin{lstlisting}[
  language=yaml,
  emph={sweep,dmin},
  emphstyle=\bfseries\color{ebsdorange!85!black}
]
# Delta from the simple example used for the selected 2B run.
# One MC output is reused by two masterpattern branches.
project_name: YNi2_2B_dmin_sweep_two_eulers

monte_carlo:
  fixed:
    ekev: 30.0
    ehistmin: 29.5
    totnum_el: 2000000000

masterpattern:
  sweep:
    dmin: [0.02, 0.05]
  bethe_parameters:
    fixed:
      c1: 100.0
      c2: 100.0
      c3: 50.0

screen_pattern:
  fixed:
    energymin: 29.5
    energymax: 30.0
    scalingmode: gam
    gammavalue: 0.33
  anglefile_generator:
    values:
      - [0.0, 20.0, 0.0]
      - [325.85, 89.60, 118.63]
\end{lstlisting}

\yamllistingneed
\subsection{Files Generated from YAML}
The YAML file is not submitted directly to EMsoft. EBSDmagus uses it to
generate the concrete files that EMsoft and Snakemake need: MC/MP/SP NML files,
angle files, Bethe-parameter files, references to \texttt{.xtal} inputs,
metadata files, validation files, and database rows. Jinja handles the rendering
step so that the researcher describes the study once, while the many
stage-specific files and internal references remain consistent.

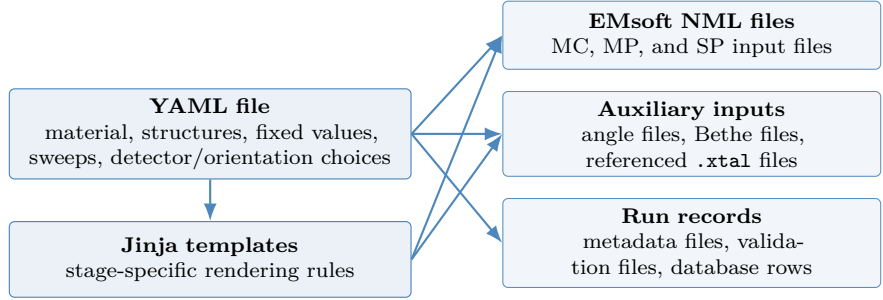
\begin{figure}[htbp]
  \centering
  \begin{tikzpicture}[
    source/.style={
      draw=ebsdblue!75,
      fill=ebsdblue!8,
      rounded corners=2pt,
      align=center,
      minimum height=1.0cm,
      text width=0.32\linewidth,
      font=\scriptsize
    },
    generated/.style={
      draw=ebsdblue!65,
      fill=ebsdblue!5,
      rounded corners=2pt,
      align=center,
      minimum height=0.9cm,
      text width=0.30\linewidth,
      font=\scriptsize
    },
    arrow/.style={-Latex, thick, draw=ebsdblue!70}
  ]
    \node[source] (yaml) {\textbf{YAML file}\\material, structures, fixed values, sweeps, detector/orientation choices};
    \node[source, below=0.55cm of yaml] (jinja) {\textbf{Jinja templates}\\stage-specific rendering rules};

    \node[generated, right=1.2cm of yaml] (aux) {\textbf{Auxiliary inputs}\\angle files, Bethe files, referenced \texttt{.xtal} files};
    \node[generated, above=0.28cm of aux] (nmls) {\textbf{EMsoft NML files}\\MC, MP, and SP input files};
    \node[generated, below=0.28cm of aux] (records) {\textbf{Run records}\\metadata files, validation files, database rows};

    \draw[arrow] (yaml) -- (jinja);
    \draw[arrow] (yaml.east) -- (nmls.west);
    \draw[arrow] (jinja.east) -- (nmls.west);
    \draw[arrow] (yaml.east) -- (aux.west);
    \draw[arrow] (jinja.east) -- (aux.west);
    \draw[arrow] (yaml.east) -- (records.west);
  \end{tikzpicture}
	  \caption{EBSDmagus renders the YAML through Jinja templates into stage-specific EMsoft NML files, angle and Bethe-parameter files, and the accompanying run records.}
  \label{fig:yaml-generated-files}
\end{figure}

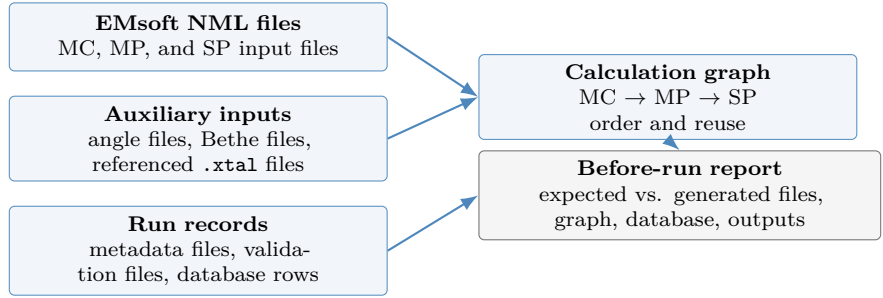
\begin{figure}[htbp]
  \centering
  \begin{tikzpicture}[
    generated/.style={
      draw=ebsdblue!65,
      fill=ebsdblue!5,
      rounded corners=2pt,
      align=center,
      minimum height=0.9cm,
      text width=0.30\linewidth,
      font=\scriptsize
    },
    checknode/.style={
      draw=black!55,
      fill=black!4,
      rounded corners=2pt,
      align=center,
      minimum height=0.9cm,
      text width=0.32\linewidth,
      font=\scriptsize
    },
    arrow/.style={-Latex, thick, draw=ebsdblue!70}
  ]
    \node[generated] (nmls) {\textbf{EMsoft NML files}\\MC, MP, and SP input files};
    \node[generated, below=0.32cm of nmls] (aux) {\textbf{Auxiliary inputs}\\angle files, Bethe files, referenced \texttt{.xtal} files};
    \node[generated, below=0.32cm of aux] (records) {\textbf{Run records}\\metadata files, validation files, database rows};
    \node[generated, right=1.2cm of aux, yshift=0.55cm] (graph) {\textbf{Calculation graph}\\MC $\rightarrow$ MP $\rightarrow$ SP\\order and reuse};
    \node[checknode, right=1.2cm of aux, yshift=-0.75cm] (preflight) {\textbf{Before-run report}\\expected vs. generated files, graph, database, outputs};

    \draw[arrow] (nmls.east) -- (graph.west);
    \draw[arrow] (aux.east) -- (graph.west);
    \draw[arrow] (records.east) -- (preflight.west);
    \draw[arrow] (graph.south) -- (preflight.north);
  \end{tikzpicture}
	  \caption{Before submission, the report compares the generated inputs and run records with the planned calculation graph.}
  \label{fig:yaml-generated-checks}
\end{figure}

Before submission, the report compares the generated NML, angle,
Bethe-parameter, and metadata files with the database and calculation graph.
\Cref{fig:yaml-generated-files} traces the rendering step, and
\cref{fig:yaml-generated-checks} shows the subsequent comparison. The file
references must agree with the database and graph for every branch in a sweep.

\clearpage
\section{Portable Database and RDM Record}
\label{app:database}

EBSDmagus stores the run database as a single SQLite file so that it
can travel with YAML inputs, reports, calculation graphs, provenance traces, and
output checks. For RDM integration, the database records what the researcher
intended, which files EBSDmagus generated, which jobs Snakemake submitted, which
jobs completed, and where the checked output files are located.

\begin{figure}[H]
  \centering
  \begin{tikzpicture}[
    tablebox/.style={
      draw=ebsdblue!65,
      fill=ebsdblue!6,
      rounded corners=2pt,
      align=left,
      minimum height=1.4cm,
      text width=0.29\linewidth,
      font=\scriptsize
    },
    satellitetable/.style={
      draw=ebsdblue!65,
      fill=ebsdblue!6,
      rounded corners=2pt,
      align=left,
      minimum height=1.1cm,
      text width=0.28\linewidth,
      font=\scriptsize
    },
    versiontable/.style={
      draw=ebsdblue!65,
      fill=ebsdblue!6,
      rounded corners=2pt,
      align=left,
      minimum height=0.9cm,
      text width=0.28\linewidth,
      font=\scriptsize
    },
    arrow/.style={-Latex, thick, draw=ebsdblue!70}
  ]
    \node[tablebox] (jobs) {\textbf{\texttt{jobs}}\\stage, project, structure, parameters and parent hashes, generated namelist path, status, SLURM id};
    \node[satellitetable, above right=0.55cm and 1.0cm of jobs] (artifacts) {\textbf{\texttt{job\_artifacts}}\\expected output paths, HDF5 files, energy files, completion markers, validation JSON};
    \node[satellitetable, right=1.0cm of jobs] (slurm) {\textbf{\texttt{job\_slurm\_context}}\\SLURM id, partition, submit command, scheduler status, timestamps};
    \node[satellitetable, below right=0.55cm and 1.0cm of jobs] (history) {\textbf{\texttt{job\_status\_history}}\\old status, new status, transition time, recovery audit trail};
    \node[versiontable, below=0.8cm of jobs] (schema) {\textbf{\texttt{schema\_version}}\\database compatibility marker};

    \draw[arrow] (jobs.east) -- (artifacts.west);
    \draw[arrow] (jobs.east) -- (slurm.west);
    \draw[arrow] (jobs.east) -- (history.west);
    \draw[arrow] (schema.north) -- (jobs.south);
  \end{tikzpicture}
	  \caption{Conceptual database schema. The \texttt{jobs} table records the planned calculations; output, scheduler, and status-history tables attach the records used for checking, recovery, provenance, and RDM integration.}
  \label{fig:database-schema-overview}
\end{figure}
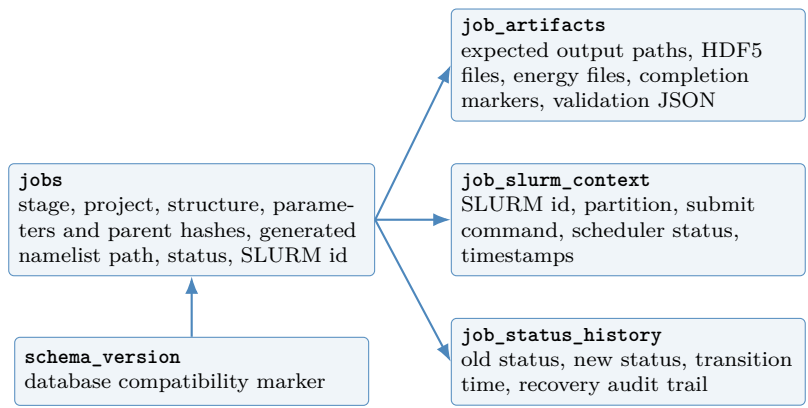

The schema centres on the \texttt{jobs} table. Other tables attach output
records, scheduler observations, status-history entries, and schema-version
information to each planned job. \Cref{fig:database-schema-overview} gives the
overall structure. \Cref{tab:dba-schema-reading-guide,tab:database-status-vocabulary}
describe its relationships and status terms, while
\cref{tab:rdm-map,tab:database-schema-overview} set out its RDM use and
individual tables. The EBSDmagus source tree contains the raw SQL schema and
migration code.

\begin{longtable}{p{0.22\linewidth}p{0.22\linewidth}p{0.44\linewidth}}
\caption{Technical reading guide for the database schema. The table summarises the implemented relationships without reproducing the full SQL definition.}
\label{tab:dba-schema-reading-guide}\\
\toprule
\textbf{Relationship or field} & \textbf{Cardinality} & \textbf{Interpretation} \\
\midrule
\endfirsthead
\toprule
\textbf{Relationship or field} & \textbf{Cardinality} & \textbf{Interpretation} \\
\midrule
\endhead
\texttt{jobs.job\_hash} &
Unique per planned job &
Stable calculation identity used to connect generated inputs, scheduler context, output checks, status history, and provenance. The integer \texttt{id} is a local row id; \texttt{job\_hash} is the cross-table key used by EBSDmagus tools. \\
\texttt{jobs} $\rightarrow$\\ {\footnotesize\texttt{job\_artifacts}} &
1:0/1 &
Rows in \texttt{job\_artifacts} are keyed by \texttt{job\_hash}. They record expected output paths and validation-file paths for a job. Missing rows indicate that the generated-job record is incomplete or legacy. \\
\texttt{jobs} $\rightarrow$\\ {\footnotesize\texttt{job\_slurm\_context}} &
1:0/1 current context &
Scheduler context stores the current or recent scheduler record for a submission attempt. It is not the full scheduler audit trail; scheduler captures and status history preserve the broader record. \\
\texttt{jobs} $\rightarrow$\\ {\footnotesize\texttt{job\_status\_history}} &
1:n &
Each status mutation can append an audit row with old status, new status, and timestamp. These rows form the recovery/reset/resubmission trail. \\
\texttt{schema\_version} &
Singleton &
The table has one compatibility row. Setup/bootstrap owns schema creation and migration; runtime tools check it read-only and fail clearly on missing, older, or newer schemas. \\
\texttt{jobs.parameters} and parent hashes &
Structured JSON fields &
\texttt{jobs.parameters} stores the stage-specific parameter set. MP and SP records include the exact MC parent hash, and SP records also include the MP parent hash. Optional JSON/text fields such as scheduler extras carry execution detail that remains attached to a job without becoming the primary relational key. \\
\bottomrule
\end{longtable}

\begin{longtable}{p{0.20\linewidth}p{0.70\linewidth}}
\caption{Status vocabulary used by the database and recovery tools. The database stores status as text; EBSDmagus commands define and interpret the vocabulary.}
\label{tab:database-status-vocabulary}\\
\toprule
\textbf{Status} & \textbf{Meaning in the run record} \\
\midrule
\endfirsthead
\toprule
\textbf{Status} & \textbf{Meaning in the run record} \\
\midrule
\endhead
\texttt{todo} &
Planned but not currently submitted; eligible for submission when dependencies are satisfied. \\
\texttt{pending} / \texttt{submitted} &
Handed to the scheduler or waiting in the scheduler queue; scheduler context should explain the current observation where available. \\
\texttt{running} &
Believed to be executing or otherwise active according to database and scheduler records. \\
\texttt{done} &
Complete after EBSDmagus has checked output records; scheduler and database records alone are not enough. \\
\texttt{failed} &
Known failed or structurally incomplete; recovery tools can reset affected work while preserving reusable upstream outputs. \\
\texttt{needs\_attention} &
Repeated or ambiguous recovery condition that should not be silently retried without inspection. \\
\bottomrule
\end{longtable}

\begin{longtable}{p{0.24\linewidth}p{0.34\linewidth}p{0.32\linewidth}}
\caption{RDM integration map linking the EBSDmagus run package to data repositories and institutional RDM systems.}
\label{tab:rdm-map}\\
\toprule
\textbf{Package element} & \textbf{What it contributes} & \textbf{RDM use} \\
\midrule
\endfirsthead
\toprule
\textbf{Package element} & \textbf{What it contributes} & \textbf{RDM use} \\
\midrule
\endhead
YAML file &
Study description: structures, stages, fixed parameters, sweeps, and detector/orientation choices. &
Human-readable record and reusable entry point for reruns or extensions. \\
Rendered Snakemake config and profiles &
Run-local database path, output/log locations, user profile, cluster profile, resource settings, and scheduler-submission choices. &
Records how the requested calculations were mapped onto a concrete local or cluster execution environment. \\
Source structure files &
CIF/POSCAR/CONTCAR inputs and selected or generated \texttt{.xtal} files where they are part of the run package. &
Connects the run record to the crystallographic source used for EMsoft input generation. \\
SQLite database &
Job rows, status history, scheduler context, output-file links, hashes, and schema version. &
Portable provenance core for institute import, archive inspection, and recovery auditing. \\
Reports and calculation graphs &
Before- and after-run findings, generated-input checks, graph structure, and status overlays. &
Records the match between planned run, submitted calculations, and completed outputs. \\
Validation files and completion markers &
Worker-written records for expected output-file checks and lightweight completion markers. &
Helps detect cases where scheduler success does not match the files visible on the filesystem. \\
Provenance traces &
Backward chain from selected final outputs to generated inputs and parent MC/MP output files. &
Allows a selected output to be traced back to its generated inputs and parent calculations. \\
Output records &
HDF5 paths, validation files, \texttt{.done} completion markers, previews, selected hashes or listings. &
Connects metadata to the computational outputs; large files may remain in the external archive. \\
Scheduler records and logs &
SLURM accounting, queue snapshots, controller logs, and worker logs for selected submitted jobs. &
Shows how cluster records were interpreted during normal execution and recovery. \\
Manifest and checksums &
Inventory of archived files, selected hashes, and source paths. &
Lets a reader or local RDM import script check that the package contents are complete. \\
\bottomrule
\end{longtable}

The run-level record supports archiving, queries, imports, and re-runs of a
simulation set. A repository or local RDM system can add its landing page, DOI,
licence, embargo rules, and institution-specific metadata.

\begin{longtable}{p{0.27\linewidth}p{0.31\linewidth}p{0.32\linewidth}}
\caption{Conceptual database-table overview. The EBSDmagus source tree contains the implementation schema and migration code.}
\label{tab:database-schema-overview}\\
\toprule
\textbf{Table} & \textbf{What it records} & \textbf{Use in recovery and RDM} \\
\midrule
\endfirsthead
\toprule
\textbf{Table} & \textbf{What it records} & \textbf{Use in recovery and RDM} \\
\midrule
\endhead
\texttt{jobs} &
Planned job: stage, structure name, parameter JSON, project name, generated namelist path, status, hashes, SLURM id. &
Links the YAML study definition to generated EMsoft inputs, scheduler submissions, reuse decisions, and later provenance queries. \\
\texttt{job\_artifacts} &
Per-job output and check paths: \texttt{.done} completion marker, data name, HDF5 output path, energy file where relevant, metadata JSON. &
Allows EBSDmagus to confirm during after-run and recovery checks that a completed job has the expected output files and validation files. \\
\texttt{job\_slurm\_context} &
Transient scheduler context: SLURM id, partition, quality of service (QoS), submit command, submission time, latest scheduler status, extra JSON. &
Separates scheduler observations from run identity and preserves enough context to explain queued, running, failed, or completed jobs. \\
\texttt{job\_status\_history} &
Status transitions with old status, new status, and timestamp. &
Provides an audit trail for recovery, reset, and resubmission decisions. \\
\texttt{schema\_version} &
Current database schema version and application timestamp. &
Makes archived databases easier to inspect later and lets runtime tools reject incompatible schemas safely. \\
\bottomrule
\end{longtable}

\clearpage
\section{Larger Calculation Example}
\label{app:large-calculation}

\Cref{sec:large-preflight} reports the 530-job parameter study prepared and
checked before submission. The compact graph below shows how the five beam
energies, two transport depths, two $d_{\min}$ values, two Bethe settings, and
SP variations produce 96,000 detector patterns. The supporting material
contains the full YAML, Graphviz \cite{Gansner2000Graphviz} graph, and
before-run report.

\begin{figure}[htbp]
  \centering
  \resizebox{0.95\linewidth}{!}{%
  \begin{tikzpicture}[
    stage/.style={
      draw=ebsdblue!65,
      fill=ebsdblue!5,
      rounded corners=3pt,
      minimum width=2.8cm,
      text width=2.5cm,
      minimum height=1.12cm,
      align=center,
      font=\footnotesize
    },
    big/.style={
      draw=black!55,
      fill=black!4,
      rounded corners=3pt,
      minimum width=3.0cm,
      text width=2.72cm,
      minimum height=1.12cm,
      align=center,
      font=\footnotesize
    },
    note/.style={
      draw=black!45,
      fill=black!4,
      rounded corners=3pt,
      minimum width=11.8cm,
      text width=11.35cm,
      minimum height=0.95cm,
      align=center,
      font=\scriptsize
    },
    arrow/.style={-Latex, thick, draw=ebsdblue!70}
  ]
    \node[stage] (yaml) {Large YAML\\checked before\\submission};
    \node[big, right=1.05cm of yaml] (mc) {10 MC\\5 energies $\times$\\2 depth limits};
    \node[big, right=1.05cm of mc] (mp) {40 MP\\2 \texttt{dmin} $\times$\\2 Bethe settings};
    \node[big, right=1.05cm of mp] (sp) {480 SP\\pattern centre and\\gamma scaling};
    \node[stage, right=1.05cm of sp] (patterns) {96,000 implied\\detector\\patterns};
    \node[note, below=0.82cm of mp] (note) {Checks: 530 EMsoft jobs, 536 graph nodes, and 1,588 edges; no duplicate output names or orphaned downstream calculations};

    \draw[arrow] (yaml) -- (mc);
    \draw[arrow] (mc) -- node[above,font=\tiny]{fanout 4} (mp);
    \draw[arrow] (mp) -- node[above,font=\tiny]{fanout 12} (sp);
    \draw[arrow] (sp) -- (patterns);
    \draw[arrow] (yaml.south) |- (note.west);
    \draw[arrow] (note.east) -| (patterns.south);
  \end{tikzpicture}%
  }
  \caption{Expansion of the larger study checked before submission. Each MC output feeds four MP calculations, and each MP output feeds twelve SP calculations. The supporting material contains the complete graph, YAML file, and before-submission report.}
  \label{fig:large-preflight-dag-appendix}
\end{figure}
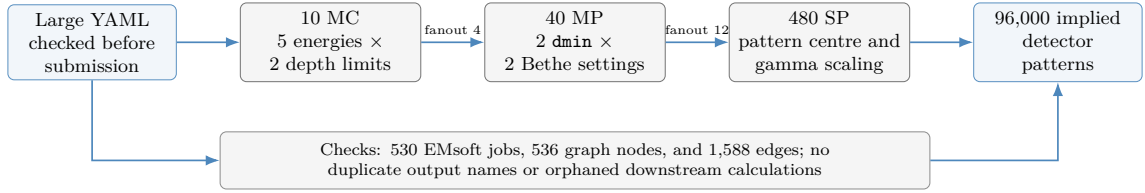

\section{Run and Recovery Details}
\label{app:run-details}

The normal and recovered runs leave database snapshots, scheduler records,
reports, and graph-status images. These records document both normal completion
and the decisions made before interrupted work resumed.

\subsection{Normal Run Checks}
\begin{figure}[H]
  \centering
  \begin{subfigure}[t]{0.82\linewidth}
    \centering
    \includegraphics[width=\linewidth]{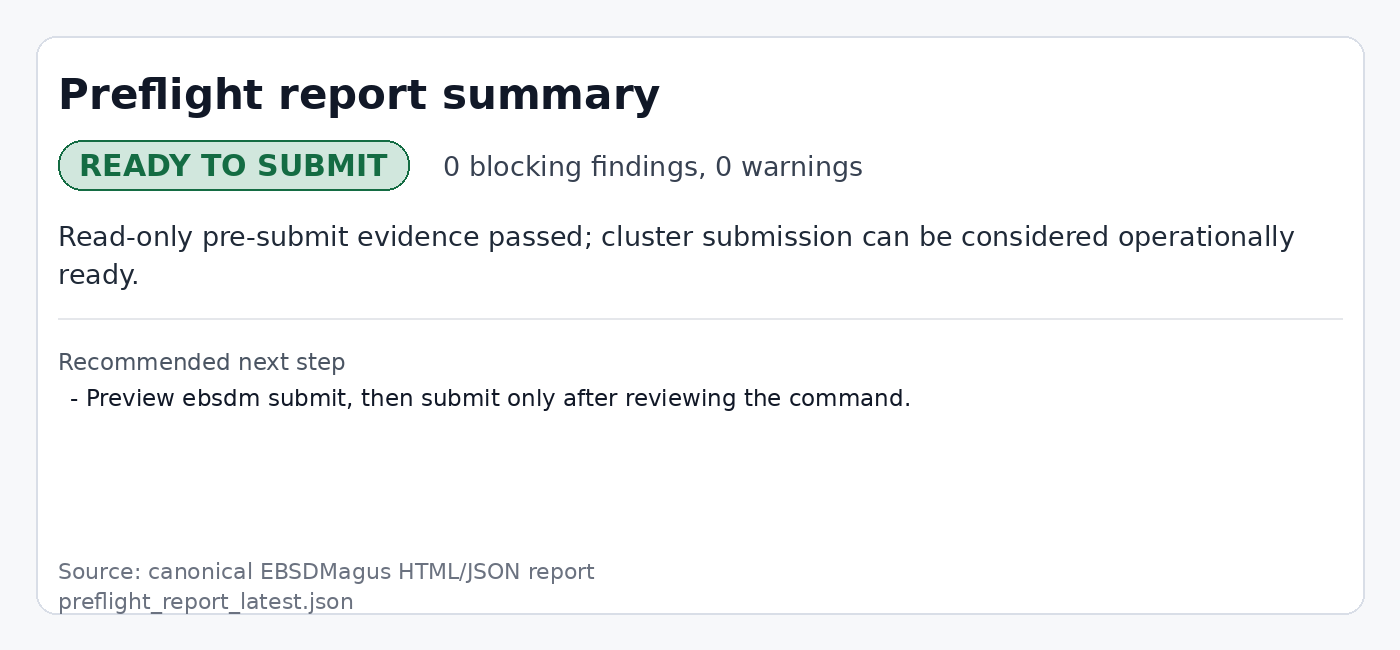}
    \caption{Summary before submission.}
  \end{subfigure}
  \par\medskip
  \begin{subfigure}[t]{0.82\linewidth}
    \centering
    \includegraphics[width=\linewidth]{reports/postflight_summary_header.png}
    \caption{Summary after completion.}
  \end{subfigure}
  \caption{Report summaries for a normal submission: (a) readiness before submission and (b) final check after completion.}
  \label{fig:normal-report-summaries-appendix}
\end{figure}

The before-run report lists the generated inputs, calculation graph, database
rows, and expected outputs before submission. The after-run report summarises
database status, scheduler records, completion markers, validation files, HDF5
output checks, and selected previews after execution.
\Cref{fig:normal-report-summaries-appendix} compares compact views before
submission and after completion; the EBSDmagus run archive retains the full
HTML and JSON reports.

\subsection{Observed Controller and Filesystem Interruptions}
\label{app:observed-interruptions}
\Cref{sec:recovery-runs} draws on two separate interruptions that occurred
during normal cluster tests. In the earlier run, the Snakemake controller was no longer running after
the MC worker had completed, but the preserved records did not reveal why the
controller ended. The corresponding database snapshots, scheduler records,
worker files, HDF5 checks, and four calculation-graph views document how
EBSDmagus retained the MC result and completed the remaining eight jobs.

The later \SI{2e9}{} electron run recorded repeated stale filesystem handles
before the controller terminated. Its separate archive contains the records
before recovery, the valid MC and MP outputs, the stale MP database entry, the
documented comparison and database update, and the final SP-only continuation.
The earlier archive records an otherwise unexplained controller loss, whereas
the later one records a shared-filesystem disruption and the resulting
disagreement between worker, scheduler, and database information.

\subsection{Controlled Worker-Cancellation Tests}
\label{app:worker-cancellation}
\Cref{sec:recovery-runs} describes the observed controller and
shared-filesystem interruptions. For a controlled comparison, we cancelled one running MP job in
two separate runs to compare Snakemake's \texttt{--keep-going} and fail-fast
behaviour. Full HTML and JSON reports remain in the EBSDmagus run archive.

\subsubsection{Worker Cancellation with Keep-Going}
\begin{figure}[H]
  \centering
  \begin{subfigure}[t]{\linewidth}
    \centering
    \includegraphics[width=\linewidth]{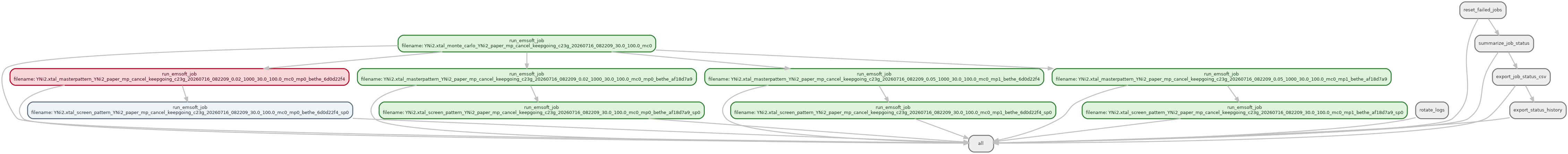}
    \caption{Before recovery: one MP branch failed, its dependent SP calculation remained unresolved, and the unaffected branches completed.}
  \end{subfigure}
  \vspace{0.4em}
  \begin{subfigure}[t]{\linewidth}
    \centering
    \includegraphics[width=\linewidth]{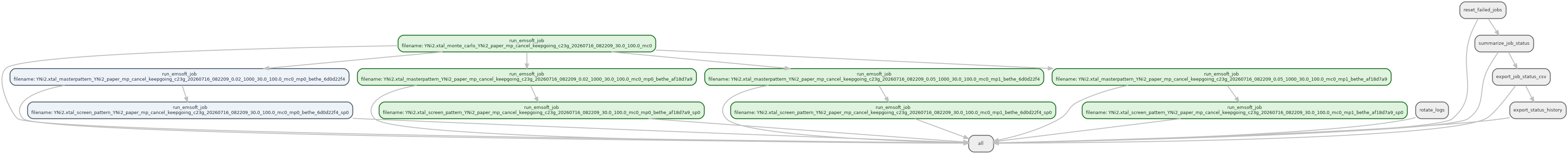}
    \caption{After targeted reset: the failed MP calculation is ready to run again.}
  \end{subfigure}
  \vspace{0.4em}
  \begin{subfigure}[t]{\linewidth}
    \centering
    \includegraphics[width=\linewidth]{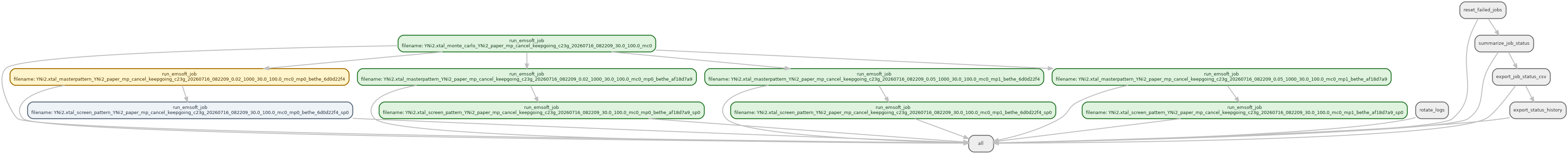}
    \caption{After resubmission: the reset MP calculation is queued and the dependent SP calculation is still waiting.}
  \end{subfigure}
  \vspace{0.4em}
  \begin{subfigure}[t]{\linewidth}
    \centering
    \includegraphics[width=\linewidth]{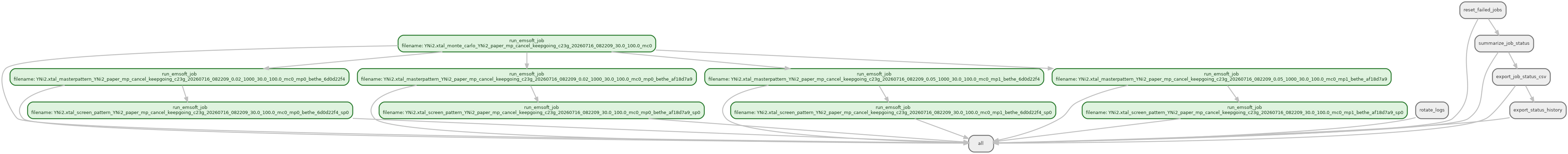}
    \caption{Final check after the replacement MP and dependent SP calculations completed.}
  \end{subfigure}
  \caption{Controlled cancellation of a master-pattern worker job with \texttt{--keep-going} enabled. (a) The failed branch remains unresolved while unaffected branches complete. (b) A targeted reset returns the failed MP calculation to the runnable state. (c) The affected branch is submitted again. (d) The final check records all selected EMsoft calculations as complete.}
  \label{fig:mp-worker-cancellation-keepgoing-appendix}
\end{figure}

We enable \texttt{--keep-going} in the current cluster profile. After one
running MP job was cancelled to simulate a worker-node failure, unaffected branches
continued; the cancelled MP branch and its dependent SP calculation remained
unresolved. EBSDmagus reset only the affected branch. After resubmission, the
replacement MP calculation and dependent SP calculation completed, and the final
after-run report classified all 9 EMsoft jobs as complete.
\Cref{fig:mp-worker-cancellation-keepgoing-appendix} follows the affected branch
from cancellation to the final completed check.

\subsubsection{Fail-Fast Worker Cancellation}
\begin{figure}[H]
  \centering
  \begin{subfigure}[t]{\linewidth}
    \centering
    \includegraphics[width=\linewidth]{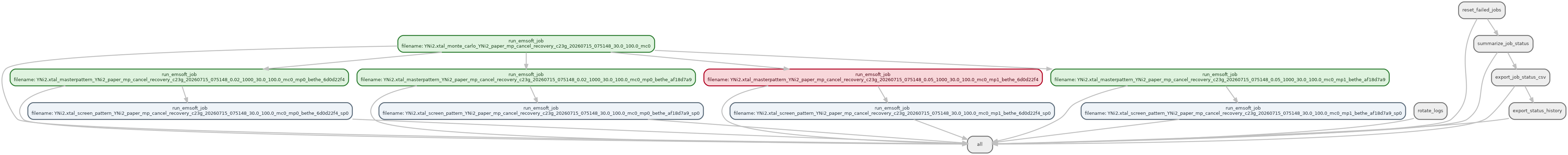}
    \caption{One cancelled MP job before recovery; the dependent SP calculation remains unscheduled in the fail-fast run.}
  \end{subfigure}
  \vspace{0.4em}
  \begin{subfigure}[t]{\linewidth}
    \centering
    \includegraphics[width=\linewidth]{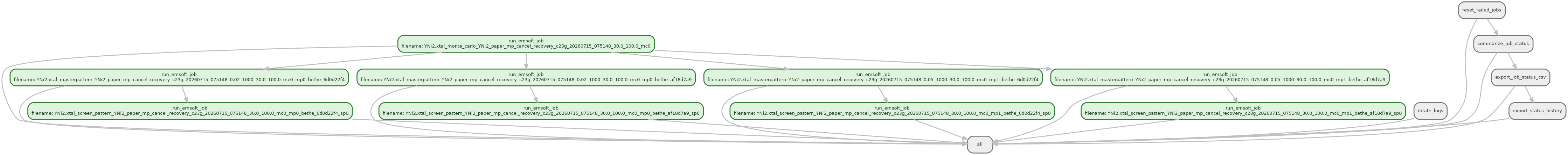}
    \caption{Recovered result after targeted reset, resubmission, output-file checks, and preview generation.}
  \end{subfigure}
  \caption{Fail-fast master-pattern worker-cancellation example. (a) After one MP worker job was cancelled, no new SP work was scheduled. (b) Recovered result after targeted reset, resubmission, output-file checks, and preview generation.}
  \label{fig:mp-worker-cancellation-failfast-appendix}
\end{figure}

In this run, Snakemake was launched without \texttt{--keep-going}. After one MP
worker job was cancelled, Snakemake stopped and did not schedule work in
otherwise independent branches. During recovery, EBSDmagus preserved completed
upstream work, reset only the affected branch, resubmitted it, and checked the
final outputs.
\Cref{fig:mp-worker-cancellation-failfast-appendix} shows the failed state and
the recovered result.

\subsection{Records Used During Recovery}
During recovery, EBSDmagus compares database entries, SLURM records, filesystem
outputs, worker-written \texttt{.done} completion markers and validation files,
HDF5 output checks, and logs because these sources operate independently and
may disagree. It then determines whether the next calculation has the inputs
needed to proceed;
researchers assess the physical content of the simulated patterns separately.
\Cref{tab:recovery-records} summarises the records and the question answered by
each one.

\begin{longtable}{p{0.24\linewidth}p{0.35\linewidth}p{0.31\linewidth}}
\caption{Records compared when EBSDmagus checks and recovers an interrupted run.}
\label{tab:recovery-records}\\
\toprule
\textbf{Record checked} & \textbf{Example question} & \textbf{Use during recovery} \\
\midrule
Database &
Which job was planned, and what status did the controller record? &
Records the planned jobs and determines what can be reconciled or resumed. \\
SLURM record &
Did the scheduler see the worker job, and how did it terminate? &
Distinguishes pending/running work from completed, cancelled, failed, or missing scheduler records. \\
\texttt{.done} completion marker &
Did the worker reach the declared completion point for this calculation? &
Lightweight worker-side completion record, but not sufficient by itself. \\
Validation file &
Did the worker validate the expected output file after writing it? &
Catches cases where scheduler success and filesystem output disagree. \\
HDF5 output &
Is the expected output file present and structurally readable? &
Confirms that dependent stages have reasonable input files, or that final SP outputs are structurally readable. \\
Logs and reports &
What did the controller and worker report before interruption? &
Preserves human-readable records for review and for narrow reset/resume decisions. \\
\bottomrule
\end{longtable}

Researchers begin recovery with a dry run and preserve the current records:
\begin{enumerate}[leftmargin=*]
  \item Preserve the database, reports, scheduler records, logs, and output-file listings before mutating anything.
  \item Run after-run/check commands to classify the observed status.
  \item Reconcile completed work when database status is stale but \texttt{.done} completion markers, validation files, output files, and scheduler records agree.
  \item Reset only the affected failed or cancelled jobs, preserving unaffected completed work and reusable upstream outputs.
  \item Resume the remaining calculations and run after-run checks again.
\end{enumerate}

\subsection{Rejected Input and Output Mismatches}
\label{app:rejection-checks}
EBSDmagus checks the file relationships and the minimum structural content that
the next calculation requires, rejecting incompatible parent calculations and
incomplete outputs before they are used by later work. The automated test suite
includes deliberately inconsistent job
records and HDF5 files for the cases summarised in
\cref{tab:rejection-checks}.

\begin{longtable}{>{\raggedright\arraybackslash}p{0.19\linewidth}>{\raggedright\arraybackslash}p{0.34\linewidth}>{\raggedright\arraybackslash}p{0.37\linewidth}}
\caption{Incompatible inputs and incomplete outputs rejected by the current EBSDmagus checks.}
\label{tab:rejection-checks}\\
\toprule
\textbf{Check} & \textbf{Rejected case} & \textbf{Reason for rejection} \\
\midrule
Parent selection &
An MP job points to an MC result whose beam energy, depth settings, structure, or other parent-defining parameters contradict the requested calculation. &
The dependency resolver reports an incompatible parent instead of selecting a file because its name appears plausible. \\
MC output &
The HDF5 file is unreadable, omits \texttt{accum\_e}, \texttt{accum\_z}, or required energy metadata, contains arrays of the wrong rank, or has energy axes inconsistent with \texttt{numEbins}. &
MP and SP calculations require the transport distributions and a consistent energy grid; file presence alone is insufficient. \\
Combined MP output &
The HDF5 file omits copied MC data or required master-pattern datasets, has inconsistent northern and southern hemisphere arrays, or disagrees internally on energy-bin count or crystal identity. &
SP generation requires a structurally consistent combined file containing both the master-pattern data and its matching MC contribution. \\
SP output &
The expected \texttt{EMData/EBSD/EBSDPatterns} dataset is absent, unreadable, non-numeric, or has fewer than two dimensions. &
The scheduler may report success even though no usable screen-pattern array was written. \\
\bottomrule
\end{longtable}

\subsection{SLURM Profiles}
\label{app:slurm-profiles}
We used SLURM through Snakemake's cluster-generic executor interface.
EBSDmagus separates the user/run
profile from the cluster profile: material and simulation choices live in YAML,
while partitions, wall times, GPU requests, accounts, logs, and worker
shell/module setup live in the cluster profile. Cluster profiles are
configurable, although we tested only the RWTH SLURM environment. Other
schedulers remain untested.

Supporting another batch system would require a scheduler-specific executor
profile, as part of the cluster profile, that maps
the EBSDmagus resource requests to that scheduler's submission command, queries
and cancels its jobs, and translates its status values for the run checks. The
study YAML, EMsoft templates, calculation relationships, and SQLite record
would remain unchanged. Any such profile would need validation against the
target cluster before production use.

Researchers edit the YAML description, from which EBSDmagus renders the EMsoft
inputs, inventories the generated files, records the planned jobs in the
database, and hands the ordered calculations to Snakemake and SLURM. After each
worker finishes, completion markers, validation files, and HDF5 output checks
give EBSDmagus information that is independent of scheduler status.

\subsection{Mock-SLURM as a Local Testing Ground}
\label{app:mock-slurm}
We developed Mock-SLURM within EBSDmagus as a lightweight local testing
environment for the scheduler boundary. It implements the subset of
\texttt{sbatch}, \texttt{squeue}, and \texttt{sacct} used by the workflow,
assigns local job identifiers, and stores each job's state in an isolated JSON
file. With the Mock-SLURM directory placed first on \texttt{PATH}, Snakemake's
cluster-generic executor and the EBSDmagus submission and status commands call
these replacements in the same way that they call the corresponding cluster
commands. Developers can check submission arguments, exported
environment variables, log paths, scheduler-status mapping, and database
transitions without using a cluster allocation or waiting in a queue.

In its default scheduler-only mode, Mock-SLURM reads the properties in a
generated jobscript, simulates a short pending--running--completed sequence, and
creates the declared outputs and scheduler logs without starting EMsoft, giving
developers a fast check of calculation submission and status handling. The
environment variable \texttt{MOCK\_SLURM\_FAIL=1} instead records a failed job
and leaves the expected outputs absent, which exercises failure and recovery
paths reproducibly.

The stronger executable mode is enabled with
\texttt{MOCK\_SLURM\_EXECUTE\_JOBSCRIPT=1}. In this mode, Mock-SLURM runs the
actual generated jobscript through its declared shell, propagates variables
passed through the SLURM \texttt{--export} option, captures standard output and
error logs, records the exit code, and reports the live state through its
\texttt{squeue} and \texttt{sacct} replacements. An expected output exists only
if the jobscript creates it, so missing outputs, invalid worker commands, and
worker-startup errors remain visible. The test covers the transfer from the
Snakemake-generated command through the submission adapter to the worker shell,
including the Python and module paths passed to the worker.

Mock-SLURM covers the scheduler boundary used in local development. It cannot
reproduce site scheduling policy, resource allocation, GPU and OpenCL
behaviour, compute-node modules, accounting, or the visibility and failure
characteristics of a shared filesystem. Even small test submissions consume
shared resources and, depending on the site's accounting policy, can affect the
priority of later jobs. Developers therefore run local dry-runs and Mock-SLURM
first, reserving cluster tests for behaviour that requires the real scheduler,
worker environment, filesystem, or accelerator hardware. A small live-cluster
smoke test follows changes to submission or worker execution, and full
validation of the MC, MP, and SP sequence remains necessary before relying on a revised cluster path
for production work.

\clearpage
\section{Archived Run Package}
\label{app:run-package}
The manuscript archive contains the selected completed run, the large
before-submission check, separate controller-loss and shared-filesystem
recovery records, both worker-cancellation tests, and a selected provenance
trace. The accompanying manifest records the exact file paths and commit
references; \cref{tab:run-package-contents} lists the retained material for
each case.

\begin{longtable}{p{0.28\linewidth}p{0.62\linewidth}}
\caption{Material retained in the manuscript run package.}
\label{tab:run-package-contents}\\
\toprule
\textbf{Package component} & \textbf{Records retained} \\
\midrule
Selected \SI{2e9}{} run &
YAML, database snapshot, before-/after-run reports, calculation graph, selected output previews, and provenance trace. \\
Large before-submission check &
YAML excerpts, generated-input records, calculation graph, pre-submission report, and summary of planned job counts and detector patterns. \\
Controller-loss recovery &
Database snapshots, scheduler records, reports, calculation-graph status images, and notes showing how completed upstream work was reconciled before continuation. \\
Shared-filesystem interruption &
Database snapshots, controller and worker logs, scheduler records, completion and validation files, HDF5 checks, recovery reports, and the SP-only continuation after the valid MC and MP outputs were retained. \\
Fail-fast worker cancellation &
Before-recovery and final recovered database snapshots, after-run reports, scheduler records, output-file list, and preview images. \\
\texttt{--keep-going} worker cancellation &
Before-recovery, reset, readiness, submit, recovery-pending, final recovered database snapshot, final after-run report, calculation-graph status images, and scheduler accounting. \\
\bottomrule
\end{longtable}

\clearpage
\section{Terminal Interface}
\label{app:tui-screenshots}
The optional terminal user interface (TUI) is implemented with Textual and runs
in an ordinary terminal. It therefore remains available over SSH on a shared
cluster without a desktop session or graphical display server. The command-line
tools remain fully usable when the optional Textual dependency is not installed.

\begin{figure}[htbp]
  \centering
  \begin{subfigure}[t]{0.48\linewidth}
    \centering
    \includegraphics[width=\linewidth]{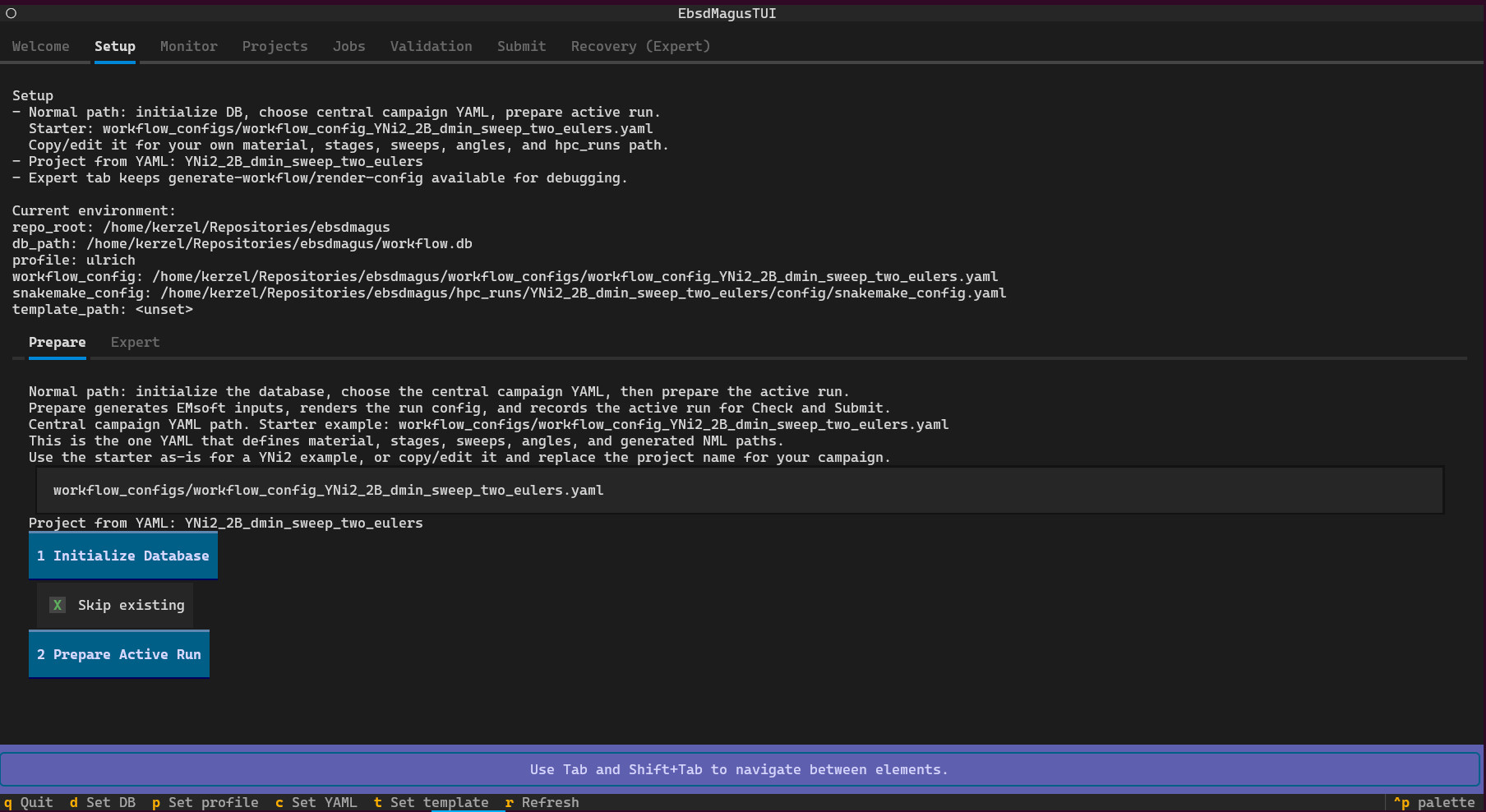}
    \caption{Setup view.}
  \end{subfigure}
  \hfill
  \begin{subfigure}[t]{0.48\linewidth}
    \centering
    \includegraphics[width=\linewidth]{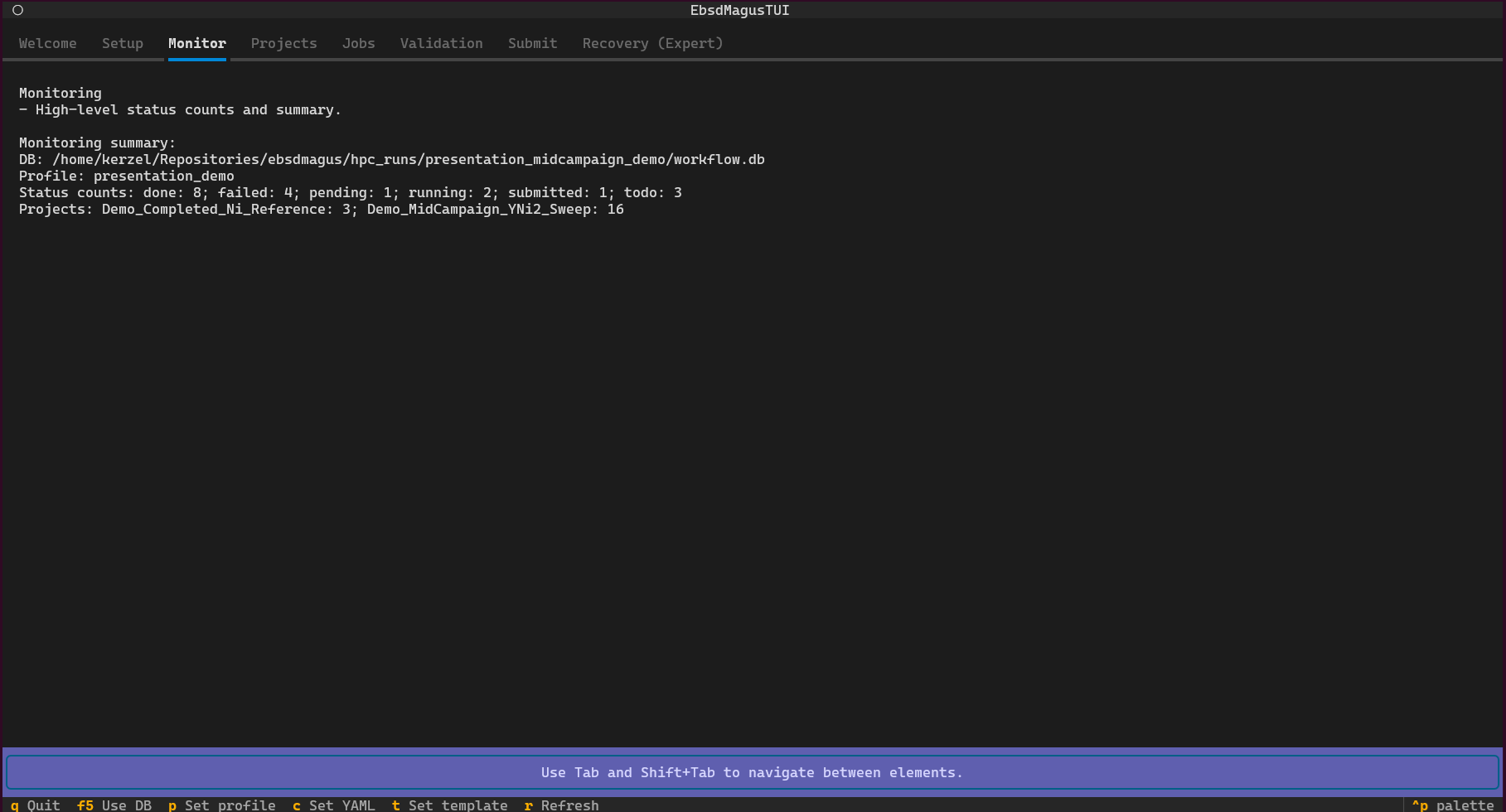}
    \caption{Monitoring view.}
  \end{subfigure}
	  \caption{Terminal user interface: (a) setup view for selecting YAML inputs and run information; (b) monitoring view for inspecting job status during or after a run.}
  \label{fig:tui-optional}
\end{figure}

The TUI presents setup, validation, submission, monitoring, job inspection, and
recovery in separate views. These views call the same EBSDmagus commands and
query the same SQLite database as the command-line interface; they do not
maintain a second execution path or a separate copy of the run record. The
setup and monitoring views in \cref{fig:tui-optional} show how users select the
YAML and database, prepare the calculation set, and inspect the recorded job
status. Recovery operations remain explicitly selected actions rather than
automatic responses to every failed job.

\clearpage
\section*{Declaration of generative AI and AI-assisted technologies in the manuscript preparation process}
The authors used OpenAI Codex for language and structural editing and
\LaTeX{} preparation, under their direction. The authors determined the
scientific content, reviewed and revised the assisted text, and accept
responsibility for the final manuscript.

\printbibliography

\end{document}